\documentclass[galaxies,article,accept,pdftex,moreauthors]{Definitions/mdpi} 
\firstpage{1} 
\pubvolume{1}
\issuenum{1}
\articlenumber{0}
\pubyear{2023}
\copyrightyear{2023}
\externaleditor{Academic Editor: Yosuke Mizuno }
\datereceived{31 May 2023} 
\daterevised{1 September 2023} 
\dateaccepted{4 September 2023} 
\datepublished{ } 
\hreflink{https://doi.org/} 

\usepackage{pdflscape}	
\usepackage{mathtools}
\usepackage{amsmath}
\usepackage{adjustbox}
\usepackage{multirow}
\usepackage{upgreek}
\usepackage{amssymb}
\usepackage{graphicx}
\usepackage{graphics}
\usepackage{multirow}
\usepackage{color}
\usepackage{xcolor}
\usepackage{bm}	
\usepackage{comment}
\usepackage{upgreek}
\usepackage{float}
\usepackage{graphicx}

\def\nat{Nature}

{\newif\ifnotend
\notendtrue
\def\veclist{ABCDEFGHIJKLMNOPQRSTUVWXYZabcdefghijklmnopqrstuvwxyz.}
\def\top#1#2.{#1}
\def\tail#1#2.{#2.}
\loop\expandafter\xdef\csname v\expandafter\top\veclist\endcsname%
{{\noexpand\bf\expandafter\top\veclist}}
\edef\veclist{\expandafter\tail\veclist}
\if\veclist.\notendfalse\fi\ifnotend\repeat}

\def\kpc{\,\mathrm{kpc}}

\def\msun{\,{\rm M}_\odot}

\def\km2s2{{\rm\,km^2\,s^{-2}}}

\def\ltsima{$\; \buildrel < \over \sim \;$}
\def\simlt{\lower.5ex\hbox{\ltsima}}
\def\gtsima{$\; \buildrel > \over \sim \;$}
\def\simgt{\lower.5ex\hbox{\gtsima}}
\newcommand{\kmskpc}{\ensuremath{\,\mathrm{{km\ s}^{-1}\ {kpc}^{-1}}}}

\setkeys{Gin}{draft=false}

\Title{Resonant Effects of a Bar on the Galactic Disk Kinematics
Perpendicular to Its Plane}

\TitleCitation{Resonant Effects of a Bar on the Galactic Disk Kinematics
Perpendicular to Its Plane}

\Author{Vladimir Korchagin 
 $^{1,}$*\orcidB{}, Artem Lutsenko $^{2}$\orcidD{} , Roman Tkachenko $^{1}$\orcidA{}, Giovanni Carraro $^{2}$\orcidC{} and Katherine Vieira $^{3}$\orcidE{}}

\AuthorNames{V. Korchagin, A. Lutsenko, R. Tkachenko, G. Carraro and K. Vieira}
\AuthorCitation{Korchagin, V.; Lutsenko, A.; Tkachenko, R.; Carraro G.; Vieira, K.}

\address{%
$^{1}$ \quad Institute of Physics, Southern Federal University, Stachki Avenue 124, Rostov-on-Don 344090, Russia; rtkachenko@sfedu.ru 
\\
$^{2}$ \quad Dipartimento di Fisica e Astronomia, Università di Padova, Vicolo Osservatorio 3, I-35122 Padova, Italy; artem.lutsenko@studenti.unipd.it (A.L.); giovanni.carraro@unipd.it (G.C.)\\
$^{3}$ \quad Instituto de Astronom\'ia y Ciencias Planetarias, Universidad de Atacama, Copayapu 485, \linebreak  Copiap\'o 1531772, Chile; katherine.vieira@uda.cl}

\corres{Correspondence: vkorchagin@sfedu.ru}

\abstract{Detailed analysis of kinematics of the Milky Way disk in the solar neighborhood based on the GAIA DR3 catalog reveals the existence 
of peculiarities in the stellar velocity distribution perpendicular to the galactic plane. {{We} 
 study the influence of resonances---the outer Lindblad resonance and the outer vertical Lindblad resonance---of a rotating bar with stellar oscillations perpendicular to the plane of the disk, and their role in shaping the spatial and the velocity distributions of stars.} We find that the $Z$ and $V_Z$ distributions of stars with respect to $L_Z$ are affected by the outer Lindblad resonance. The existence of bar resonance with stellar oscillations perpendicular to the plane of the disk is demonstrated for a long (large semi-axis 5 kpc)
 and fast rotating bar with $\Omega_{\rm b}= 60.0~\kmskpc$. 
We show also that, in the model with the long and fast rotating bar, 
some stars in the 2:1 OLR region deviate far from their original places, entering the bar region. A combination of resonance excitation of stellar motions at the 2:1 OLR region together with strong interaction of the stars with the bar potential leads to the formation of the group of 'escapees',  
i.e., stars that deviate in $R$ and $Z$---directions at large distances from the resonance region.
Simulations, however, do not demonstrate any noticeable effect on $V_Z$-distribution of stars in the 
solar neighborhood.}


\begin{document}


\section{Introduction}

Since the HIPPARCOS era it has been known that the solar neighborhood manifests complex dynamical structures such as streams and moving groups. 
The Gaia mission~\citep{Perryman,Lindeg,Fabricius,Evanss}, with its accurate observational data, shed new light the on properties of kinematical and 
structural peculiarities, both in the solar neighborhood and beyond. Examples of such structures are the {\it ridges} 
(diagonal distributions 
in the $(R, V_{\phi})$ plane) and {\it arches}, phase spirals in the $(Z, V_{Z})$ plane, bimodality in $(V_{R}, V_{\phi})$ plane including the 
Hercules stream~\citep{Antoja2018,Kawata2018}, and vertical waves~\citep{Khanna}. 

Many of the abovementioned kinematical and structural features are related to the resonant phenomena occurring within the Galactic disk. 
Resonances cause the radial migration of the stars that in turn affect the disk's density distribution and its kinematical properties. \citet{SB09} distinguished 
two types of radial migration caused by resonance phenomena. In the first type, the angular momentum of a star changes, so the star's orbit 
moves inwards or outwards, depending on whether the angular momentum is lost or gained. This type of resonance occurring in stellar disks was studied 
by \citet{SB} and \citet{Ros}, who 
showed that the resonant interaction of stars with a spiral density wave churns the stars around corotation causing their re-distribution within 
the  Galactic disk. As was stressed by \citet{SB}, such behavior has profound consequences for the chemical evolution of galaxies. Similar 'churning' 
of matter within corotation resonance in gaseous gravitating disks was shown by \linebreak  \citet{LK} who found that an unstable spiral density wave depletes mass 
around the corotation, leading eventually to an amplitude saturation of an unstable spiral. If a star-periodic gravitational field interaction 
increases a star's epicycle amplitude without changing its angular momentum, the star contributes to the density over a wider range of radii. 
\citet{SB09}, in modification of the terminology introduced by \linebreak  \citet{SB}, called theses changes in the epicycle amplitude 'blurring'.

\citet{Dehnen} demonstrated that a fast rotating bar with corotation resonance located at 3.5--5 kpc causes 
a bimodal distribution of stars in the $(V_{R}, V_{\phi})$ plane at the outer Lindblad resonance (OLR).
This idea was further confirmed by \citet{Antoja14} and \linebreak  \citet{Monari17c} who explained the bimodality in $(V_{R}, V_{\phi})$ as a result of the bar interaction with the stars at the 2:1 OLR. 

\citet{Hunt} discussed another possibility to account for the observed kinematical peculiarity in the solar neighborhood---the Hercules stream, by the 4:1 OLR resonance with a slowly rotating bar. \citet{HuntSp} demonstrated that the Hercules stream and the observed ridges in the $(R, V_{\phi})$ plane can be explained by transient winding spiral arms, independently or in combination with a long and slow bar. It is also known that ridges can be produced by a barred potential and its resonances~\citep{Antoja2018}.

\citet{Antoja2018} demonstrated recently the existence of a phase-spiral pattern seen in the solar neighborhood in the $(Z,V_Z)$---plane. The plausible explanation of the appearance of the Gaia phase-space spiral is a passage through the Milky Way disk of the Sagittarius dwarf galaxy about 
400 $\pm$ 150 Myr ago (see \citet{Antoja2018,BS18} and \linebreak  \citet{TG22}). Alternatively, the observed feature
can be explained by disk  perturbations from the subhalos~\citep{CWD18}, a perturbation generated by a buckling bar~\citep{Khop}, and a misaligned gas accretion~\citep{Khachaturyants}. In addition to the influence of resonances on dynamic structures and phase-space mixing~\citep{Antoja2018,HuntSp,Trick}, the influence of the resonance effects on chemical evolution of the Galactic disk is also well known~\citep{Mish1,Mish2,Mish3,Kchem}.

Resonances of an external perturbing potential with the stellar motions in the direction perpendicular 
to the Galactic plane were first considered by \citet{Bin81}. These resonances, however, were not studied 
in much detail. \citet{Combes90} demonstrated that vILR-inner resonance of the bar with the vertical motions 
of the stars can be responsible for the formation of the bar's peanut shape. Further, the influence of vertical 
inner Lindblad resonances on the formation of X- or peanut-shaped Galactic bulges was discussed by \citet{Quillen2002} and \citet{QUI}. 
The resonant thickening of self-gravitating disks was analytically studied by \citet{Fouvry}.
Recently, \citet{Vieira} presented a detailed analysis of 
kinematics of the Milky Way disk in the solar neighborhood using the GAIA DR3 catalogue. 
The authors found an excess of stars in the $V_{Z}$-distribution of red giants in the solar neighborhood 
at  $-40 < V_{Z} <-20~\mbox{km~s}^{-1}$ and a dearth of stars in the $V_{Z}$ distribution 
at $25 < V_{Z} <50~\mbox{km~s}^{-1}$. 

The discovery of dips and bumps in the $V_Z$-velocity distribution of stars in 
the solar neighborhood (\citet{Vieira}) posed the question if the development of such features 
can be caused by the resonance of the rotating bar with the stellar motions perpendicular to the galactic disk in the 
solar neighborhood.
The resonances of a rotating bar with stellar motions in the plane of the disk lead to 
a considerable reshuffling of matter within the plane of the disk and can also influence the 
kinematical and density distributions of stars in a Galactic disk in the direction perpendicular 
to the disk's plane.

We study in this paper how the resonances of the stellar motions with a rotating bar, occurring in 
the solar neighborhood, namely, 2:1 outer Lindblad resonance, \linebreak  1:1 outer Lindblad resonance, 
and the vertical outer Lindblad resonance can influence the disk's kinematical properties, in particular, 
the peculiar kinematical features reported by \linebreak  \citet{Vieira}.

Section~\ref{sec:method} describes the method we chose to study the resonance dynamics of the thin disk stars 
under non-axisymmetric time-dependent bar potentials. Section~\ref{sec:res} presents the results of our study. 
Section~\ref{sec:conclusion} gives a brief summary of our results.

\section{The Model}\label{sec:method} 
We study the influence of the bar on the dynamics of a thin collisionless disk, aiming at possible detection of resonance effects, both 
in the plane of the disk and perpendicular to disk's plane. The behaviour of a thin collisionless disk is simulated with help of  
the astrophysical 
package
\texttt{Galpy}~\citep{galpy}, by integrating the particle dynamics using Dormand--Prince method (\texttt{dop853}) 
\citep{SODE}---the eighth order Runge--Kutta family method with the adaptive timesteps. To generate the equilibrium collisionless disk, 
we followed the dynamics of the initial velocity and density distribution of particles in the Milky Way-like potential during three Gyr until 
the equilibrium distribution of stars is achieved. 
We then study the dynamics of the resulting equilibrium configuration for another three Gyr, while considering the influence of the bar potential. We assume that 
the bar smoothly switches on over a period of one Gyr after the axisymmetric equilibrium of the disk is settled. 
We find that the integration time of three Gyr is sufficient to study the influence of the bar---further increase of integration time does not
lead to the  qualitative change of disk dynamics. 

\subsection{Initial Distribution}

As it was mentioned above, we discuss the dynamics of a thin Galactic disk, taking into account the possible influence of a bar. 
We are interested in the influence of resonances with a bar outside the bar region; therefore, the stellar spatial distribution 
was generated outside 
four kpc using $10^6$ particles and assuming the commonly accepted exponential density distribution law $\propto\exp\{ -r/r_d \}$, where $r_d$ is 
the radial scale length of the Galactic thin disk. There is a fairly large spread of the measured values of $r_d$ ranging from two to four kpc 
\citep{DrimRd,GerRd,PiffRd,KhrapovRd}. Following \citet{Juric}, we choose in this paper the value of $r_d=2.5$ kpc. For the vertical 
distribution of particles, we use the exponential law 
$\propto\exp\{ -z/h_z \}$ with the adopted value for vertical scale height of $h_z=300$ pc~\citep{Juric,KordoRz}.
Average velocities of the particles in the radial direction and in the direction perpendicular to the disk
were chosen to be zero ${<V_R>}=0$ and ${<V_Z>}=0~\mbox{km~s}^{-1}$. To generate mean velocities of 
the particles in the azimuthal direction ${<V_{\phi}>}$, we use the disk's rotation curve in the axisymmetric Galactic potential 
\texttt{McMillan17}~\citep{McMillan} (from now on MC17), described in detail in next subsection. We adopt the values of 
the velocity dispersions of particles in the solar neighborhood $\sigma_R, \sigma_{\phi}, \sigma_Z$ equal to 31, 20 and 11 \mbox{km s}$^{-1}$ 
\citet{Vieira}. 
The dependence of the velocity dispersion of particles on the radius was adopted in the form 
$\sigma_R=\sigma_{R,0}/\exp(r/7.4\,[\kpc])$, taken from \citet{KhrapovRd,Triede} where $\sigma_{R,0}=93~\mbox{km~s}^{-1}$
to satisfy the observed value of the velocity dispersion in the solar neighborhood of $\sigma_R=31~\mbox{km~s}^{-1}$. 
The azimuthal velocity dispersion was set using the equilibrium  condition for the collisionless disk~\citep{KhrapovRd}:

\begin{equation}\label{eq:crcfKappa}
    \sigma_\varphi = \sigma_R\, \frac{\varkappa}{2\Omega} \,,
\end{equation}
where $\Omega$ and $\varkappa$ are the angular velocity and the epicyclic frequency of the disk calculated from the equilibrium rotation curve  
in the MC17 potential. For setting the velocity dispersion perpendicular to the disk direction, 
we use the ration $\sigma_Z/\sigma_R=0.36$ to satisfy the observed value of 
$\sigma_Z$ velocity dispersion in the solar neighborhood $\sigma_Z=11~\mbox{km}~\mbox{s}^{-1}$.

\subsection{Galactic Potential Models}
We adopt the axisymmetric model of the Galactic potential from~\citep{McMillan}, which includes potentials of the bulge, 
the thin and the thick disks, the gravitational potentials of the gaseous ($H_1$ and $H_2$) disks, and the gravitational 
potential of the dark matter halo. 


To model the time-dependent potential of a rotating bar, 
we add to the MC17 potential model a potential of a smoothly growing for one Gyr bar, using  the prescription taken from \citet{Dehnen}:
\clearpage
\begin{equation}\label{eq:growth}
  A(t)=\left(\frac{3}{16}\xi^5-\frac{5}{8}\xi^3+\frac{15}{16}\xi+\frac{1}{2}\right),\quad 
\end{equation}
where $\xi=2(\frac{t-t_{form}}{t_{steady}})-1$, when $t_{form}\leq t \leq t_{form}+t_{steady}$; $\xi=-1$, when 
$t\leq t_{form}$ and $\xi=1$, when  $t\geq t_{form}+t_{steady}$. 
The amplitude multiplier function $A(t)$ 
smoothly 
changes from 0 to 1 for one Gyr, so the bar potential is smoothly turned on after the equilibrium axisymmetric distribution
is settled.  

The bar potential is modeled by the equation taken from  \citet{Long}:
\begin{equation}\label{eq:bar}
\Phi_r(x,y,z)=\frac{GM_b}{2a}\ln\left(  \frac{x-a+T_-}{x+a+T_+} \right).
\end{equation}

Here, $T_{\pm}=[(a\pm x)^2 + y^2 +(b+\sqrt{c^2+z^2})^2) ]^{1/2}$.  $a,b,c$ are characteristic bar parameters and $x,y,z$ 
are Cartesian coordinates.  We adopt values of bar semi-axes equal to \linebreak  $b=1$ kpc, $c=0.3$ kpc. For the major semi-axis of the bar, $a$,  
we adopt two values: \linebreak  $a=5$ kpc---the long bar, (LB) and $a=3.5$ kpc---the short bar (SB). In accordance with \citet{Port17} 
and \citet{Kent}, we choose the bar mass equal to $M_\mathrm{b}=1.88\times 10^{10}\,\msun$, which is also in a agreement 
with the estimate of \citet{Zhao}. 

A number of studies have been undertaken to measure the bar angular \linebreak  velocity~\citep{Debat,Sanders,Clarke,Shen,Sormani}. 
\citet{Debat} estimated the angular velocity of the Galactic bar  to be $\Omega_{\rm b}= 59\pm5 \kmskpc$. Recent data 
based on Gaia proper motion measurements allowed for a more accurate estimation of the pattern speed of the Galactic bar. However, there is still 
disagreement between the results. For example, \citet{Sanders} used the VVV Infrared Astrometric Catalogue (VIRAC) 
and Gaia DR2 proper motions and estimated the angular velocity of the bar to be $\Omega_{\rm b}= 41\pm3 \kmskpc$, 
\citet{Clarke} obtained the value of $\Omega_{\rm b}= 37.5 \kmskpc$~\citep{Shen}.  \citet{Dehnen} suggested that, to explain 
some kinematical features observed in the solar neighborhood, a bar with the angular velocity of  $\Omega_{\rm b}= 55.5~\kmskpc$ 
is required. The fast bar is also supported by some other \linebreak  studies~\citep{Gardner,Antoja14,Chakra,Bissantz} 
with the bar angular velocity varying within  $\Omega_{\rm b}= \mbox{55--60}~\kmskpc$. We use in this paper two values 
of the angular velocity of the bar: $\Omega_{\rm b}$ = 40~\kmskpc \linebreak  (slow bar) and $60~\kmskpc$ (fast bar). 
Thus, we explore two values of the angular velocity of a bar and the two values of the bar's major semi-axis. 
We use the following designations for our models. The model with the fast long bar ($\Omega_{\rm b}= 60 \kmskpc$, \linebreak  a = 5 kpc) 
is designated as  ``LB60'', and the slow short bar ($\Omega_{\rm b}= 40 \kmskpc$, \linebreak  a = 3.5 kpc) is designated as ``SB40''. 
The models we use are listed in Table~\ref{tab1}. 

\begin{table}[H] 
\caption{List of different bar models.\label{tab1}}
\newcolumntype{C}{>{\centering\arraybackslash}X}
\begin{tabularx}{\textwidth}{CCC}
\toprule
\textbf{Name}	& \boldmath{$\Omega_{\rm b}~[\kmskpc]$}	& \boldmath{$a~[\mbox{\textbf{kpc}}]$}\\
\midrule
SB40 
		& 40			& 3.0\\
SB60		& 60			& 3.0\\
LB40		& 40			& 5.0\\
LB60		& 60			& 5.0\\
\bottomrule
\end{tabularx}
\noindent{\footnotesize{\textsuperscript{} Here $\Omega_{\rm b}$ is angular bar velocity and $a$ is the bar major semi-axis.}}
\end{table}
\subsection{Rotation Curves}
\begin{figure}[H]
\includegraphics[width=11 cm]{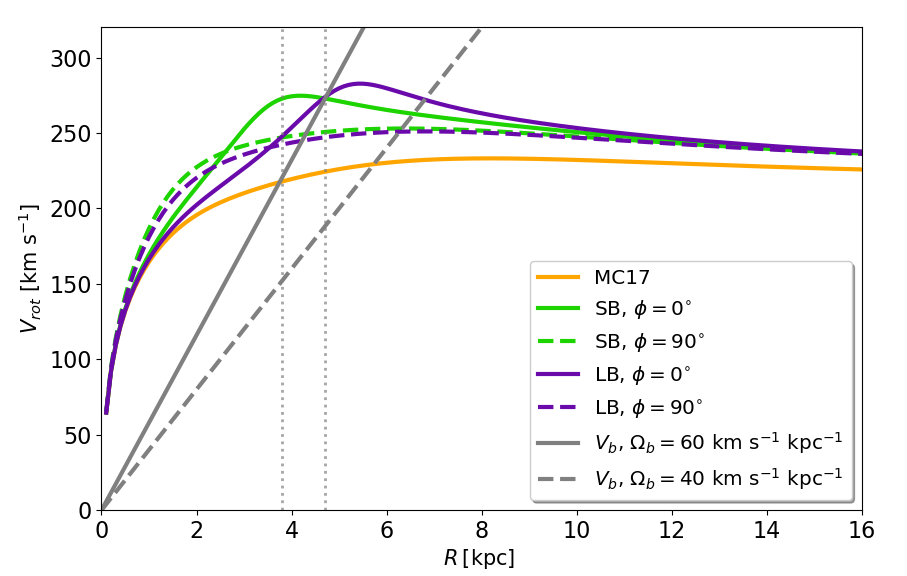}
\caption{Rotation 
 curves of axisymmetric (orange line) and non-axisymmetric models. For the models with bars, the rotation curves are shown for different directions with respect to the bar major semi-axis. Short bar---green lines. Long bar---purple lines. Grey lines---the linear rotation curves corresponding to the angular velocities of a bar. The vertical lines mark the positions of corotation in MC17 model and in the models with the bar.
\label{figRC}} 
\end{figure}

The rotation curves of our barred, and non-barred models are shown in Figure~\ref{figRC}. In the  
bar potential, the rotational curve is systematically
higher compared to that in the axisymmetric model MC17 due to the fact that a bar with mass $M_\mathrm{b}=1.88\times 10^{10}\,\msun$ is added
to the axisymmetric MC17 model. The rotation curves were determined from the equation $v_{rot}=(r d\Phi /dr)^{1/2}$, where $\Phi$ represents 
the total potential in the plane Z = 0. With the bar potential the disk rotation is not circular in its central regions, so the ``rotation curve'' depends 
now on the direction relative to the bar major axis. Figure~\ref{figRC} presents the rotation curves calculated along a few directions relative to the 
bar major axis.
As one can see from Figure~\ref{figRC}, with the adopted mass of the bar of 1.88 $\times$ 10$^{10}$  M$_{\odot}$, which is comparable to that
of the unperturbed axisymmetric disk,  the bar changes the disk's rotation, which shifts the position of corotation resonance from $\approx$ 3.8 
kiloparsecs in the MC17 model to about 5 kps.  
The long bar model satisfies the criterion of \citet{Cont}, for which $\mathbb{R}=R_{CR}/R_{bar}\geq1$. We also notice that the position of
corotation is important for studying the particle dynamics at corotation resonance, but is not of much importance in our study focused on disk dynamics at 
the outer and the vertical outer Lindblad resonances.

The galactocentric distance of the Sun is chosen to be $R_{\odot}=8.122$ kpc \linebreak  (\citet{Gravity}).
\vspace{-6pt}

\subsection{Resonances}

\subsubsection{Resonances in the Plane of the Disk}

Resonances in the Galactic plane of the disk are determined by the condition:
\begin{equation}\label{eq:reson}
   m(\Omega_{b}-\Omega)=  l\kappa\,,
\end{equation}
where $\kappa$ is the epicyclic frequency, $l$ and $m$ are integer numbers equal to $(l=0,~\pm1,~\pm2,~...)$ and $(m=1,2,~...)$. 
We use the following notation for resonances: if $(m,l)=~(2, -1),~(2, 1),\\~(1,1)$, this determines the  2:1 inner Lindblad resonance (2:1 ILR) and 2:1 outer Lindblad resonance (2:1 OLR) and 1:1 outer Lindblad resonance (1:1 OLR). We consider 
the dominating m = 2 and m = 1 modes in the bar potential, neglecting higher order modes. 
It should be noticed, however, that m = 4 Fourier component of the bar potential can influence the disk dynamics as shown by \citet{Hunt}. The corotation resonance (CR) is determined by the conditions $l=0$, when the rotation frequency of the stars is equal to the bar pattern frequency.

\begin{figure}[H]
\includegraphics[width=12.0 cm]{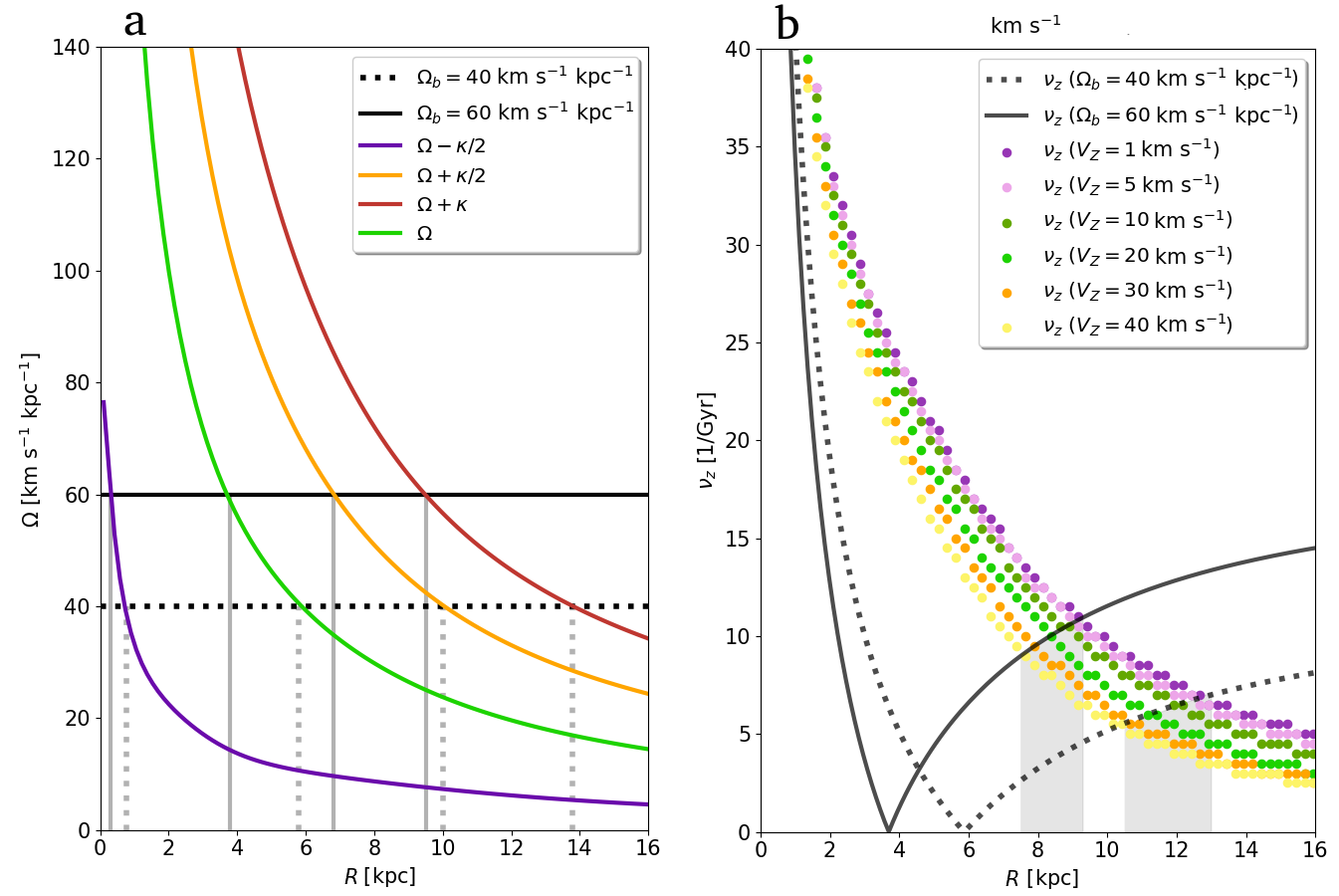}
\caption{The positions of resonances in the plane of the disk (\textbf{a}) and in perpendicular to the disk direction (\textbf{b}). 
(\textbf{a}) Black solid and dotted horizontal lines are the angular velocities of the bar.
Purple, green and orange lines show the radial dependence  of the quantities $\Omega-\kappa/2$, $\Omega$, $\Omega+\kappa/2$ 
and  $\Omega+\kappa$, respectively. The vertical lines mark the positions of 2:1 ILR, CR, 2:1 OLR and 1:1 
OLR resonances. 
(\textbf{b}) Frequencies of oscillations of particles (colored dots) perpendicular to the disk as a function of galactocentric radius. Black solid and dotted lines show the radial dependence of frequency of vertical oscillations, $\nu_{Z}$ in the left hand side of the equation~\ref{eq:vertreson}. 
Crossings correspond to the positions of the 2:1 vertical outer Lindblad resonance (2:1 vOLR) in the disk.  \label{figres}}
\end{figure}

Figure~\ref{figres} shows the positions of ILR, OLR and CR resonances in the plane of the disk in our models for bar pattern speeds $\Omega_{\rm b}= 40 \kmskpc$ and 60 $\kmskpc$. 

Our focus is on the study of the resonance effects in the solar neighborhood and beyond,  
so we exclude from consideration the disk dynamics in the central regions of the disk---namely, the inner Linblad resonance and corotation. 

\subsubsection{Vertical Resonances}

Resonance between oscillation of a star in the direction perpendicular to the plane of the disk  and rotating perturbing potential 
that has m-fold rotational symmetry occurs when the following condition is met~\citep{Bin81,Sellw,SW}:
\begin{equation}\label{eq:vertreson}
   m(\Omega_{b}-\Omega)=  l\Omega_{Z} \,,
\end{equation}
where \emph{l} is an integer number.

We will be interested in resonance condition when $l=+1$  because, for larger values of $l$, resonance conditions are achieved far from the solar 
neighborhood. The value of $m$ we assume to be equal to $m=2$. We study thus an influence of the vertical outer 2:1 Lindblad resonance (2:1 vOLR). $\Omega_{Z}$ is the frequency of the vertical oscillations, which we will use further in the form 
of $\nu_{Z}=\Omega_{Z}/2\pi$, where $\nu_{Z}$ is the linear frequency of oscillations.

{In the direction perpendicular to the disk, its density and potential are changing rapidly so the periods of oscillations 
of the particles depend on their velocities in the Z-direction at $Z=0$ and have to be determined separately.
To find the positions of the vertical resonance, we measure the frequencies of the vertical oscillations of particles  
by direct integration of the orbits of particles in the axisymmetric potential.} To do this, we generate at different radii of the disk the particles that have velocities perpendicular to the disk's plane
equal to $V_{Z}=1, 5, 10, 20, 30, 40$ $~\mbox{km~s}^{-1}$, 
and the azimuthal velocities corresponding to the equilibrium rotational velocity at a given radius. We also assume that
$V_{R}=0~\mbox{km~s}^{-1}$.
{To estimate the frequencies, we followed the dynamics of the particles during a fixed period of time and calculated the number of intersections of the particles with the mid-plane of the disk.} Results of simulations, covering the range of $V_{Z}$-velocities in the solar neighborhood, are shown in Figure~\ref{figres}b. We notice that frequencies of particle oscillations 
are in agreement with the results of \citet{Antoja2018}. \linebreak  {\citet{Beraldo} used the python package \texttt{naif} for the frequency analysis  of orbits.} In this paper, we determined the frequencies of the non-linear oscillations from the numerical simulations that we deem sufficient for our purpose.

Figure~\ref{figres}b shows the positions of the vertical resonances in our mock Galactic disk, determined in this way. 
Black solid and dotted lines demonstrate the radial dependence of the  theoretical frequency $\nu_{Z}$ from Equation~(\ref{eq:vertreson}) 
as a function of radius for two values of the bar angular velocity $\Omega_{\rm b}= 40 \kmskpc$ and 60 $\kmskpc$.
The positions of the vertical resonances are marked in Figure~\ref{figres}b by the grey zones.

As one can see from Figure~\ref{figres}b, the resonance of stellar motions with a bar that has an angular velocity 
of $\Omega_{\rm b}= 40 \kmskpc$ 
occurs at $\approx 11-13$ kpc, while resonance with the fast rotating bar that has  $\Omega_{\rm b}= 60 \kmskpc$ occurs at $\approx \mbox{8--9}$ kpc. 

\begin{figure}[H]
\includegraphics[width=.8\textwidth]{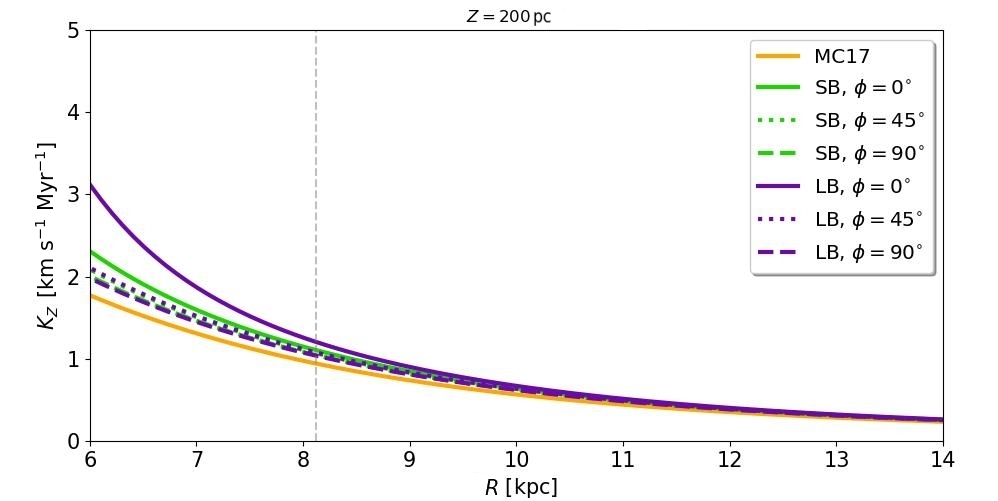}
\includegraphics[width=.8\textwidth]{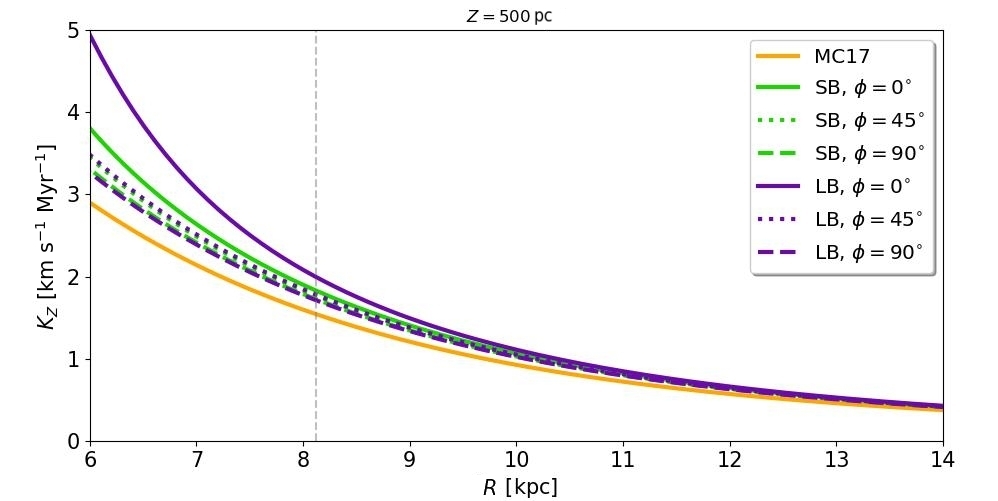}
\caption{Radial 
 dependence of the vertical component of the force for the axisymmetric MC17 potential (orange) 
and potentials with the bar with major semi-axis a = 3.5 kpc (green) and a = 5 kpc (purple). For models with the bar, the 
force dependence is shown for different angles with respect to the bar major axis. 
The vertical gray line marks the position of the Sun.  \label{figZForce}} 
\end{figure}

When analyzing the resonance regions, it is important to know variations of the vertical force $K_Z$ in the models with 
the Galactic bar $K_Z=-{\partial \Phi}/{\partial Z}$. Figure~\ref{figZForce} shows the radial dependence of the
Z-component of the gravitational force taken at Z = 200 pc and \linebreak  Z = 500 pc for the axisymmetric MC17 potential and for two models with 
bar major semi-axis a = 3.5 (SB) and a = 5  (LB) kpc.
With the bar potential, the calculations are carried out in three different directions with respect to the large semi-axis: 
$0^{\circ}, 45^{\circ} $ and $ 90^{\circ}$. {The amplification of oscillations of the particles perpendicular to the plane of the disk 
is caused by 
the periodic force from the non-axisymmetric rotating bar potential.}  The efficiency of such amplification of particle oscillations 
is determined by the difference of the bar gravitational force acting perpendicular to the disk in the direction of the bar's major and minor axes.  

As one can see from the Figure~\ref{figZForce}, for the long bar (LB) model the variation of the force at the R = 9 kpc is about 10\%, while 
at R = 13kpc the force variation is about 4\%. 
For the short bar (SB), variation of gravitational force caused by the bar is about 3\% at the \linebreak  R = 9 kpc,  and at R = 13 kpc. 

\section{Results and Discussion}\label{sec:res}

\subsection{Equilibrium in the Axisymmetric Potential}

\begin{figure}[H]
\includegraphics[width=.45\textwidth]{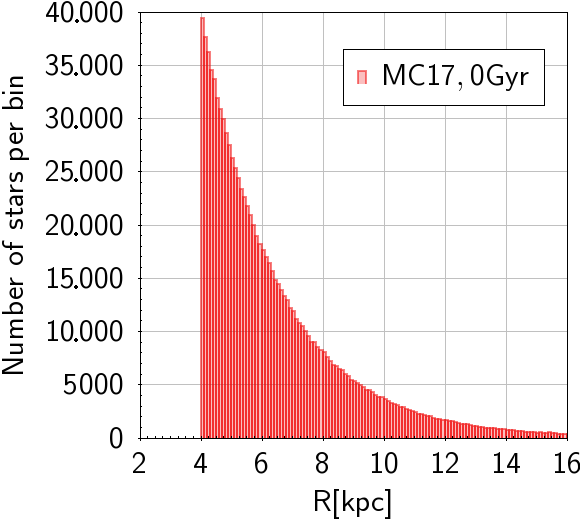}
\includegraphics[width=.45\textwidth]{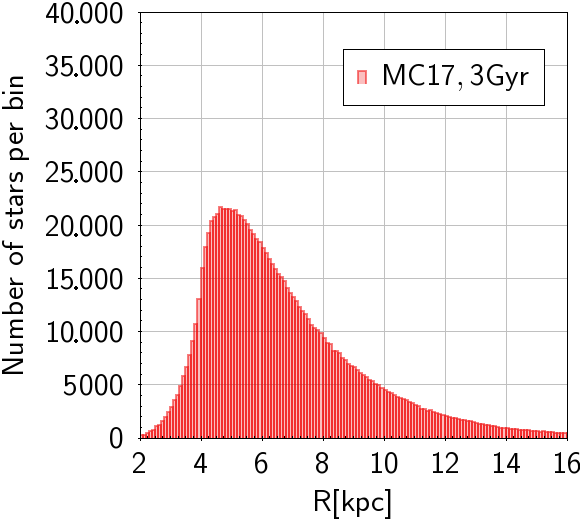}
\caption{Evolution 
 of the particle distribution in the MC17 model from the initial distribution (\textbf{left}) to the equilibrium one after three Gyr of evolution. The size of the radial bins is 0.1 kpc.}    \label{figUNBARrad} 
\end{figure}

To construct the axisymmetric equilibrium distribution in the MC17 potential model, we integrate the dynamics of the initially generated distribution of particles for five Gyr.
Figure~\ref{figUNBARrad} shows the result of such integration.  
After $\approx$ three Gyr of integration the disk rebuilds itself and reaches equilibrium, remaining unchanged during further integration, 
with the radial scale length of the disk beyond four kpc approximately equal to its initial value of $r_d=2.5~\mbox{kpc}$.

\begin{figure}[H]
\begin{adjustwidth}{-\extralength}{0cm}
\centering
\includegraphics[width=4.1cm]{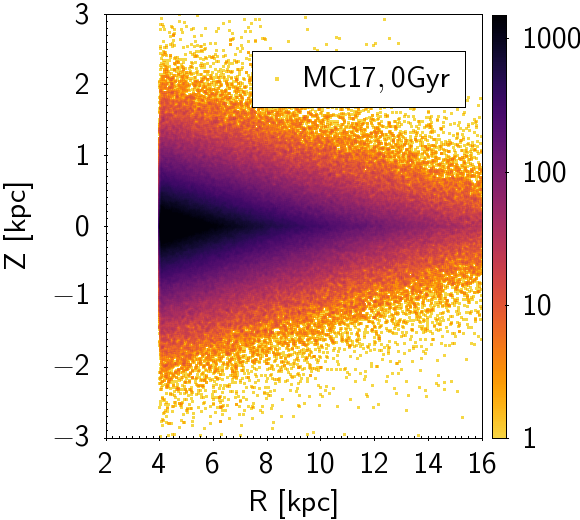}
\includegraphics[width=4.1cm]{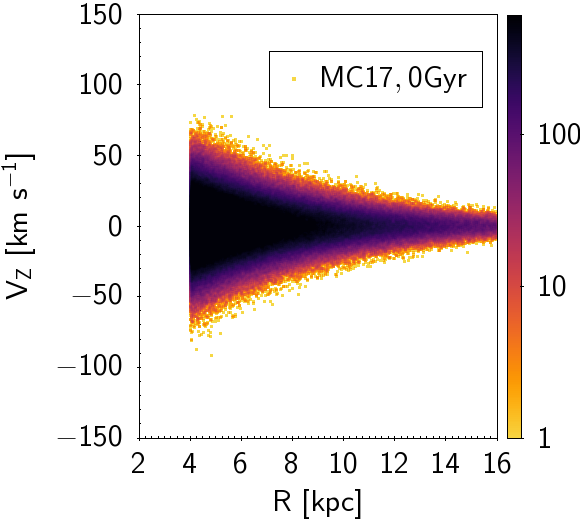}
\includegraphics[width=4.1cm]{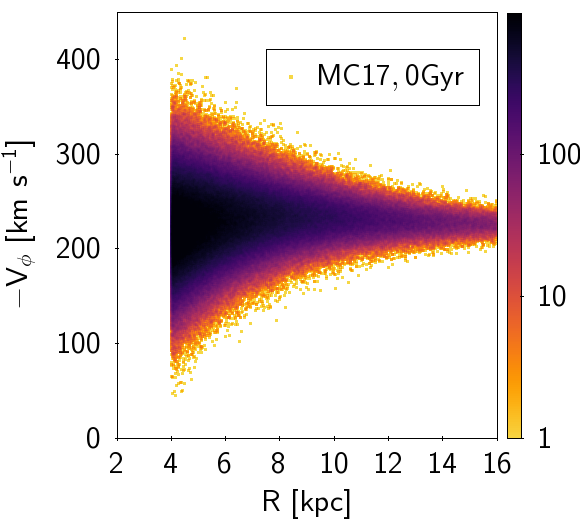}
\includegraphics[width=4.1cm]{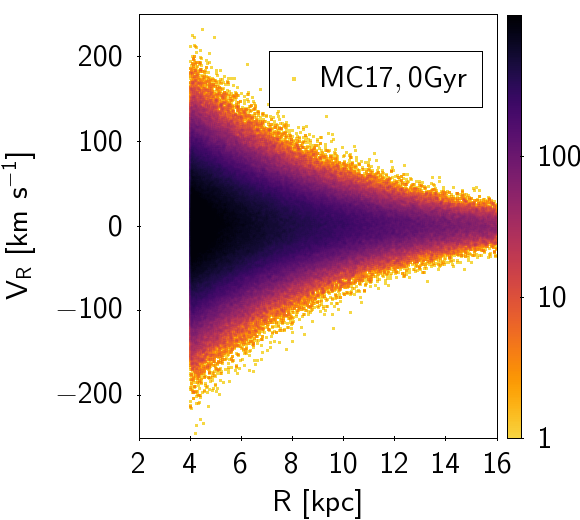}
\includegraphics[width=4.1cm]{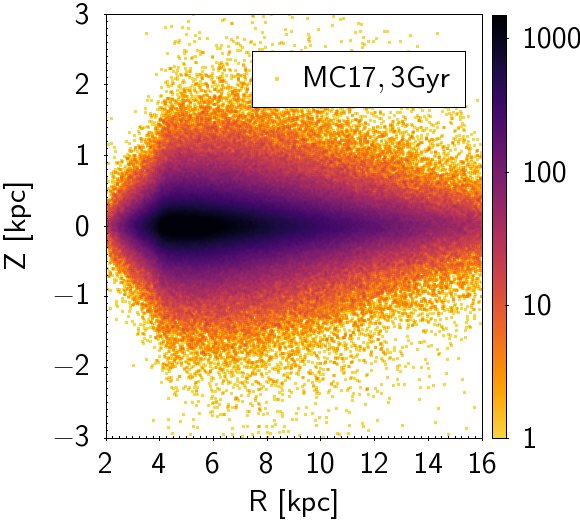}
\includegraphics[width=4.1cm]{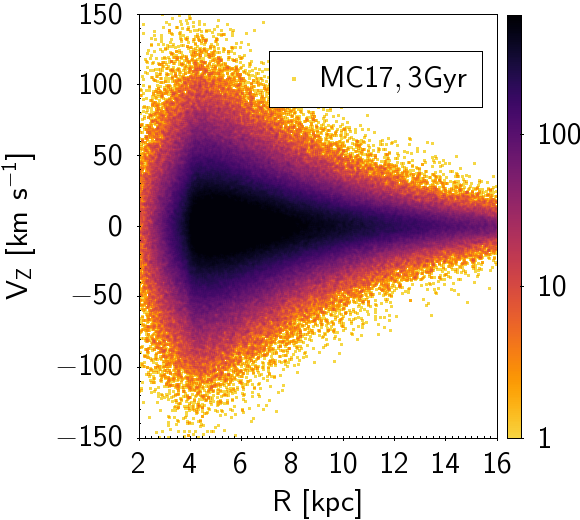}
\includegraphics[width=4.1cm]{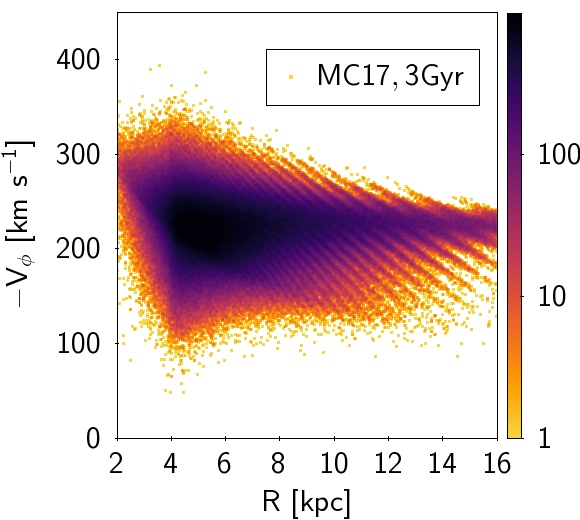}
\includegraphics[width=4.1cm]{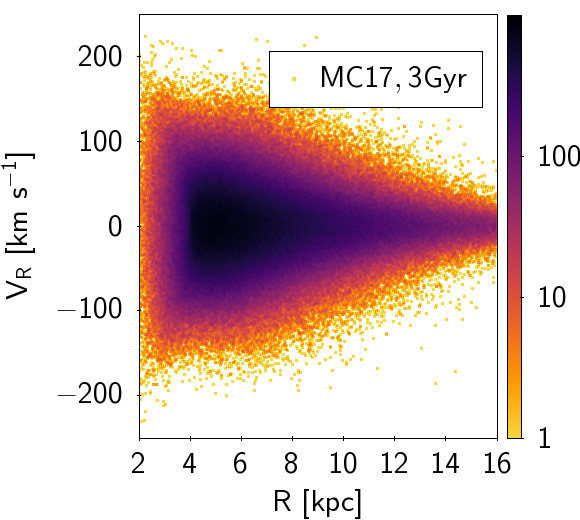}

\end{adjustwidth}
\caption{Rearrangement 
 of the initially generated distribution in the MC17 potential (top frames) after three Gyr of evolution (bottom frames). First column---the distribution of stars along the radius, second, third, and fourth columns distributions of $V_{Z}$, $V_{\phi}$ and $V_{R}$ velocity components along the radius. Particles' density distributions are encoded in the colour bars.   \label{figUNBAR4}} 
\end{figure}

Figure~\ref{figUNBAR4} shows the distributions along the radius of the particles Z-coordinate (first column), velocity component perpendicular to the disk $V_{Z}$ (second column), azimuthal velocity $V_{\phi}$ (third column), and the radial velocity of the particles $V_{R}$ (fourth column) of the initially generated distribution of particles 
(first row) and  the equilibrium distribution reached after three Gyr integration in MC17 potential (second row). 
There is a redistribution of particles both in the coordinate and in the velocity space. As one can see, 
the re-arrangement of the azimuthal velocity of the particles $V_{\phi}$ is noticeable. After three Gyr, there is seen the appearance of thin diagonal ridges. 
These ridges, seen also in simulations by \linebreak  \citet{Antoja2018} and \citet{Fux,Minchev1,Gomez}, 
result from the phase-space mixing in the plane of the disk that has initial distribution out of equilibrium~\citep{Antoja2018}. 
The equilibrium distribution established in the axisymmetric potential after three Gyr of integration was taken as the 
initial distribution to simulate the dynamics of the disk dynamics in the models with a bar.

\subsection{Dynamics in a Barred Potential}



\subsubsection{Resonances in the Plane of the Disk}

\begin{figure}[H]
\includegraphics[width=.5\textwidth]{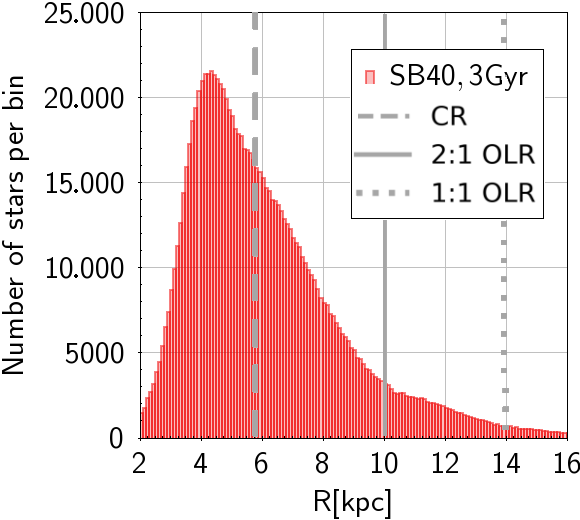}
\includegraphics[width=.5\textwidth]{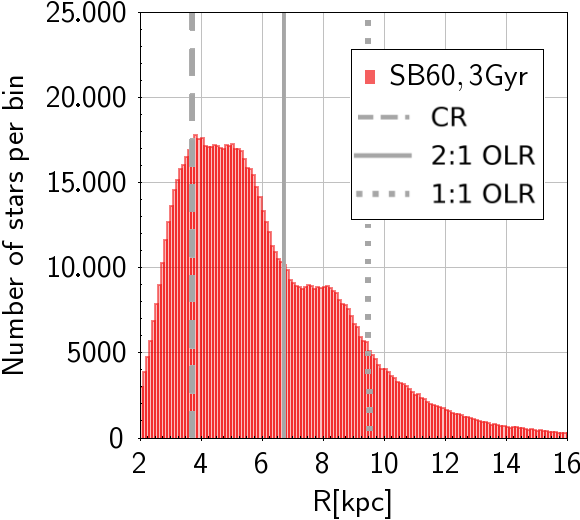}
\includegraphics[width=.5\textwidth]{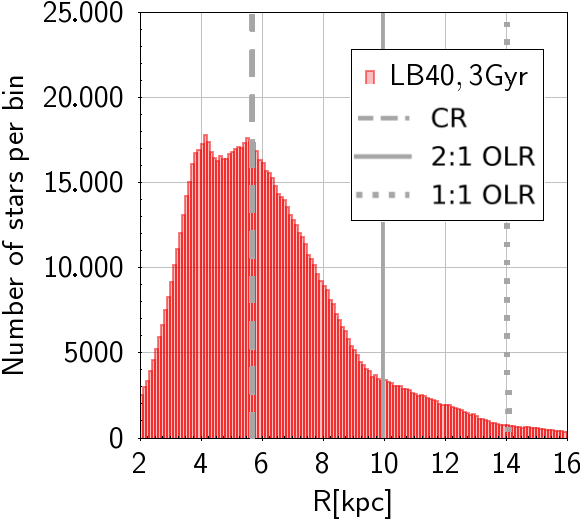}
\includegraphics[width=.5\textwidth]{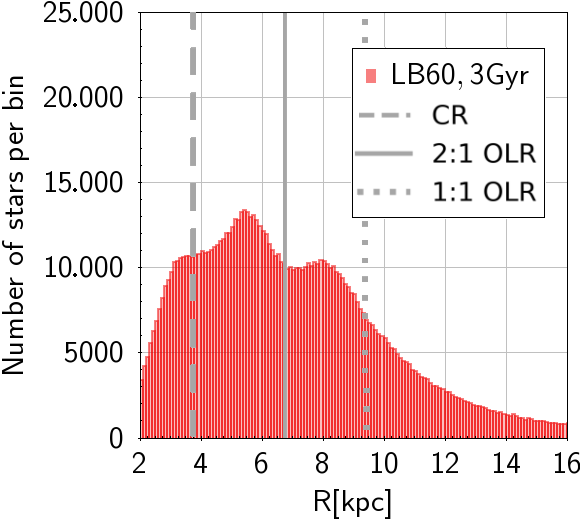}
\caption{Radial distributions of particles in four barred potentials after three Gyr of evolution \linebreak  (SB40 top left; SB60 top right; LB40 bottom left; LB60 bottom right). The size of the radial bins is 0.1 kpc. The vertical dashed, solid and dotted lines show the positions of CR, 2:1 OLR and 1:1 OLR, respectively. \label{figBARrad}} 
\end{figure}

Figure~\ref{figBARrad} demonstrates distribution of stars along the radius in four models with bar potential after three Gyr of evolution from the equilibrium state. The positions of corotation, OLR and 1:1 resonances are marked on the figure with the vertical lines. One can see the correlation of the dips on the distribution of particles along the radius with the position of the outer Lindblad resonance, especially noticeable in the models with the fast rotating bar.
This result is known from previous studies (see, e.g., \citet{Melnik}). Stars near the OLR resonance change their angular momenta, which affects the orbits, eventually causing the radial migration of the stars. The process can move the stars radially by kiloparsecs~\citep{Mikkola,SB}.

\vspace{-12pt}

\begin{figure}[H]
\begin{adjustwidth}{-\extralength}{0cm}
\centering
\includegraphics[width=4.4cm]{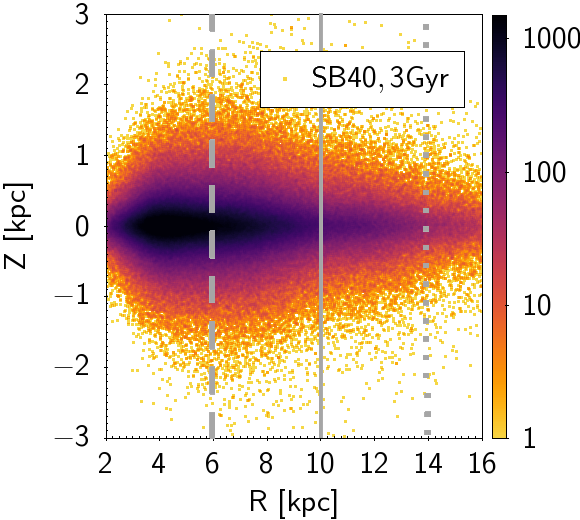}
\includegraphics[width=4.4cm]{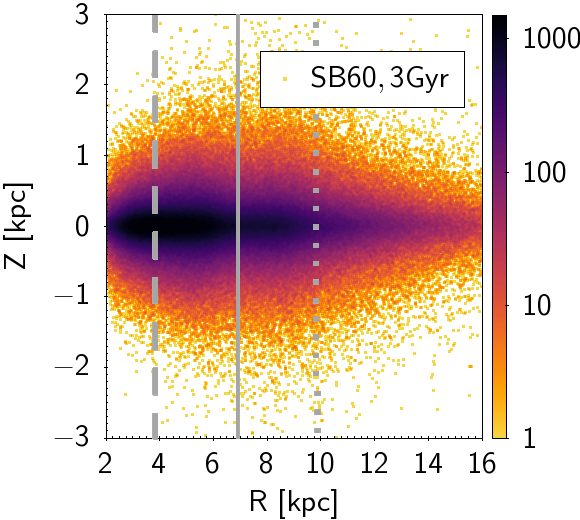}
\includegraphics[width=4.4cm]{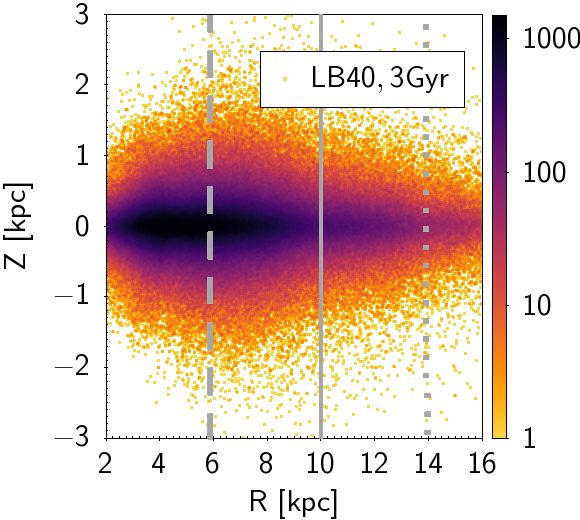}
\includegraphics[width=4.4cm]{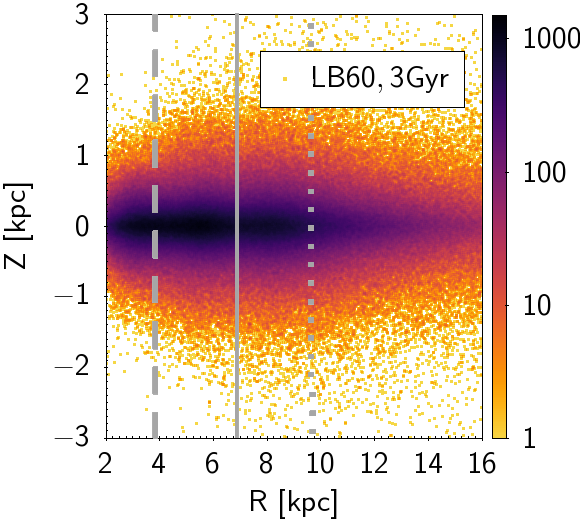}
\includegraphics[width=4.4cm]{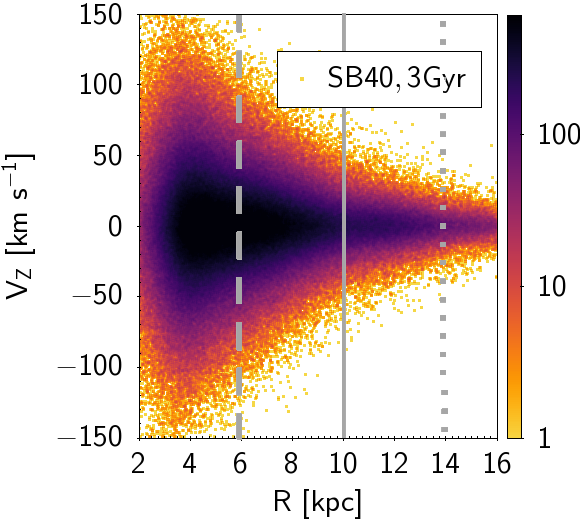}
\includegraphics[width=4.4cm]{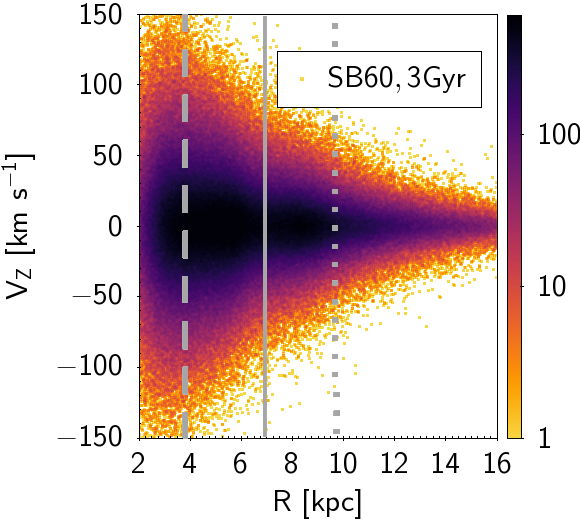}
\includegraphics[width=4.4cm]{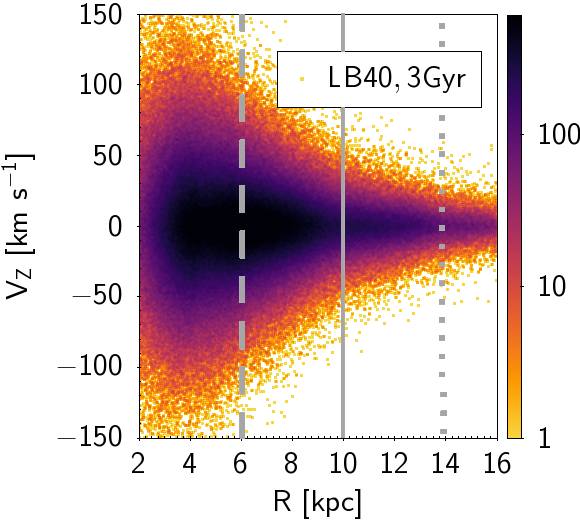}
\includegraphics[width=4.4cm]{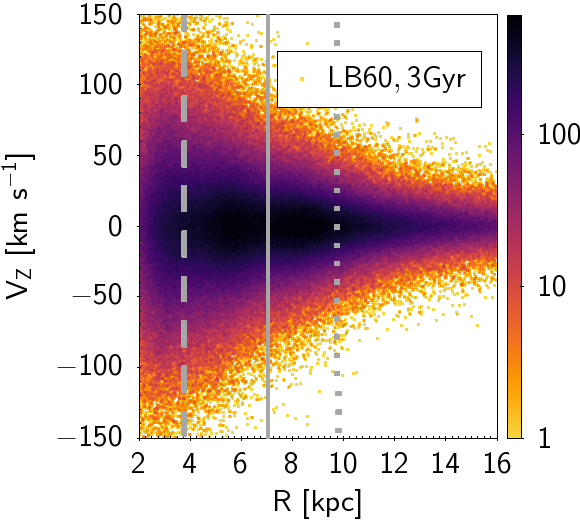}
\includegraphics[width=4.4cm]{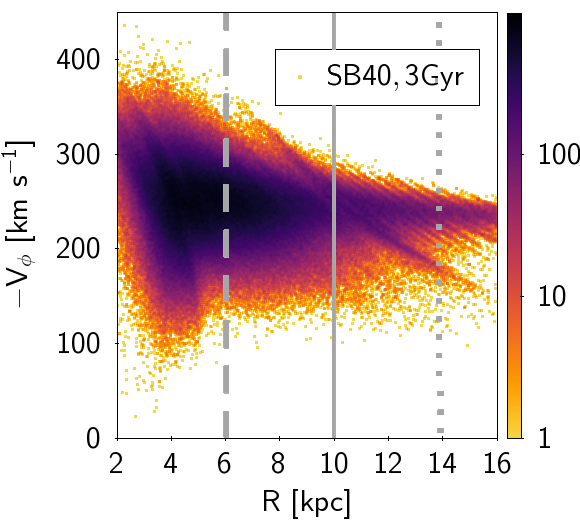}
\includegraphics[width=4.4cm]{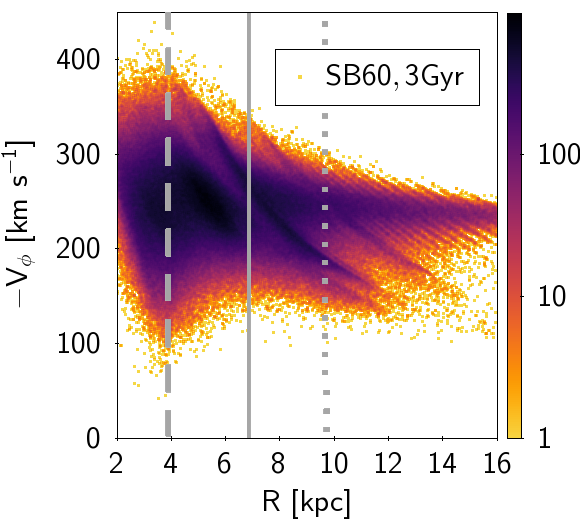}
\includegraphics[width=4.4cm]{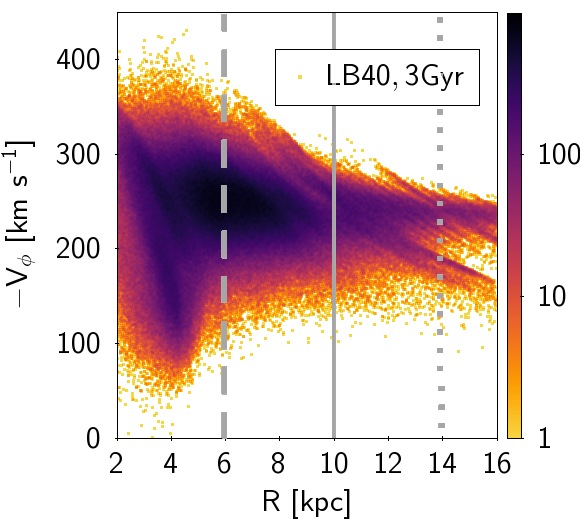}
\includegraphics[width=4.4cm]{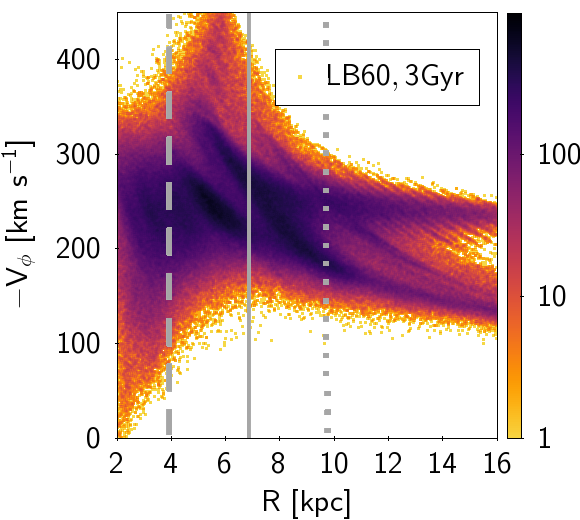}
\end{adjustwidth}
\caption{ Distribution of particles 
 shown on $(R,Z)$---plane (first row), $(R,V_Z)$---plane (second row), and $(R,V_{\phi})$---plane (third row) after three Gyr of evolution in models SB40 (first column), SB60 (second column), LB40 (third column), and LB60 (fourth column). The vertical dashed, solid and dotted lines show positions of CR, 2:1 OLR and 1:1 OLR, respectively.\label{figHorRes}}
\end{figure}

Figure~\ref{figHorRes} shows the distribution of stars in $(R,Z)$, $(R,V_Z)$ and $(R,V_{\phi})$ planes after three Gyr of orbits evolution 
in the barred potential. Diagrams $(R,Z)$, $(R,V_Z)$ do not show a noticeable effect of the bar potential on the distribution of the particles 
at OLR resonances. 
{The role of the 2:1 outer Lindblad resonance is revealed, however, as a decreasing of density of particles at OLR resonance in models with
the fast rotating bar ($\Omega_b = 60 \kmskpc $) due to proximity in these models of resonance to the bar region as shown in Figure~\ref{figBARrad}.}
The influence of the OLR resonances on the distribution of particles is also illustrated in the 
$(R,V_{\phi})$-diagram of Figure~\ref{figHorRes}, which shows the appearance of noticeable diagonal ridges seen on the figure. 
{Similar ridges, confirmed observationally~\citep{Kawata2018}, are connected to the resonant orbital structure in the barred potential that forms 
regions in the phase space with stable and unstable orbits, and hence the regions with overdensities and gaps~\citep{Michtchenko,Antoja2018,Hunt19}.}

\begin{figure}[H]

\includegraphics[width=.38\textwidth]{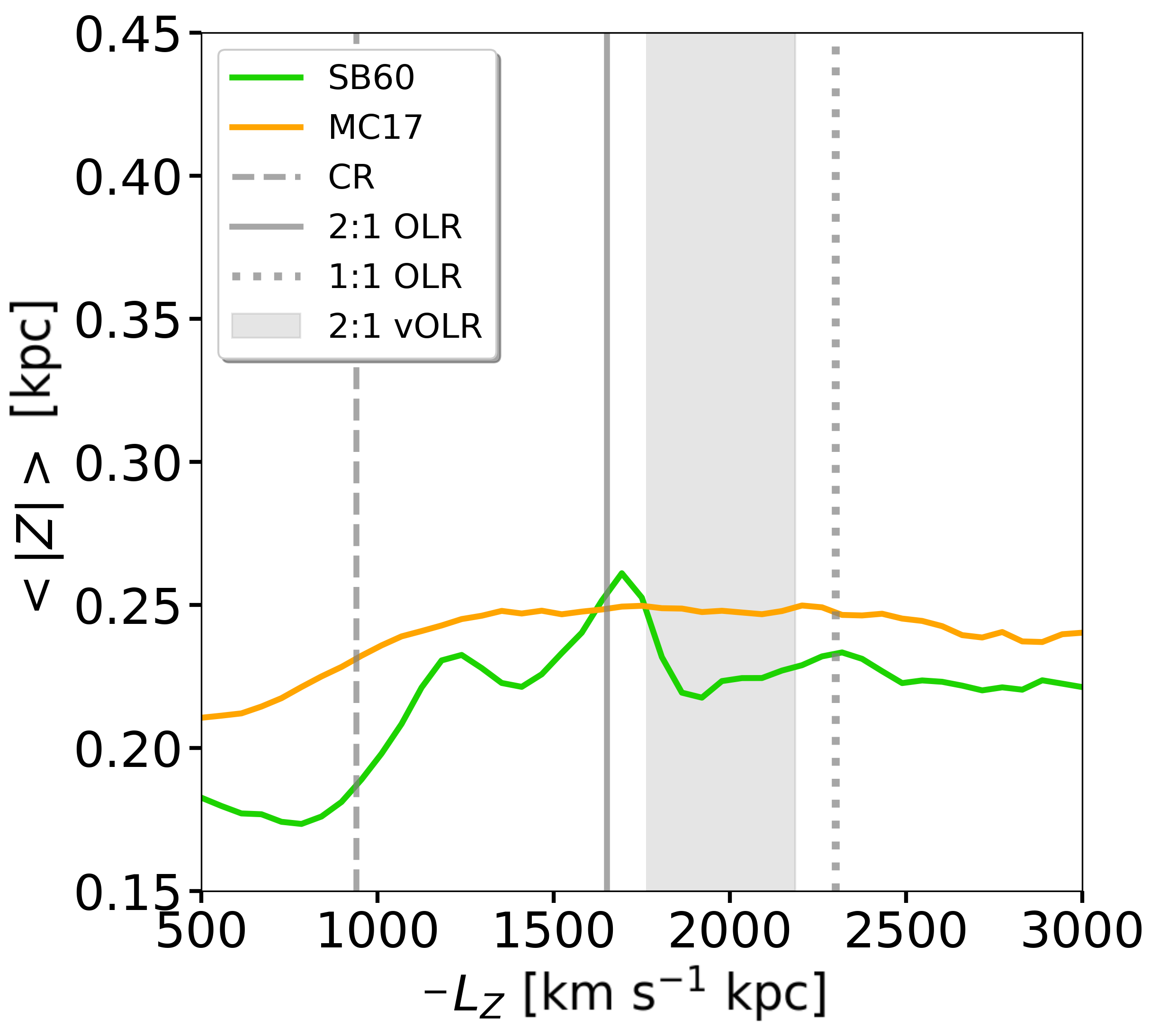}
\includegraphics[width=.37\textwidth]{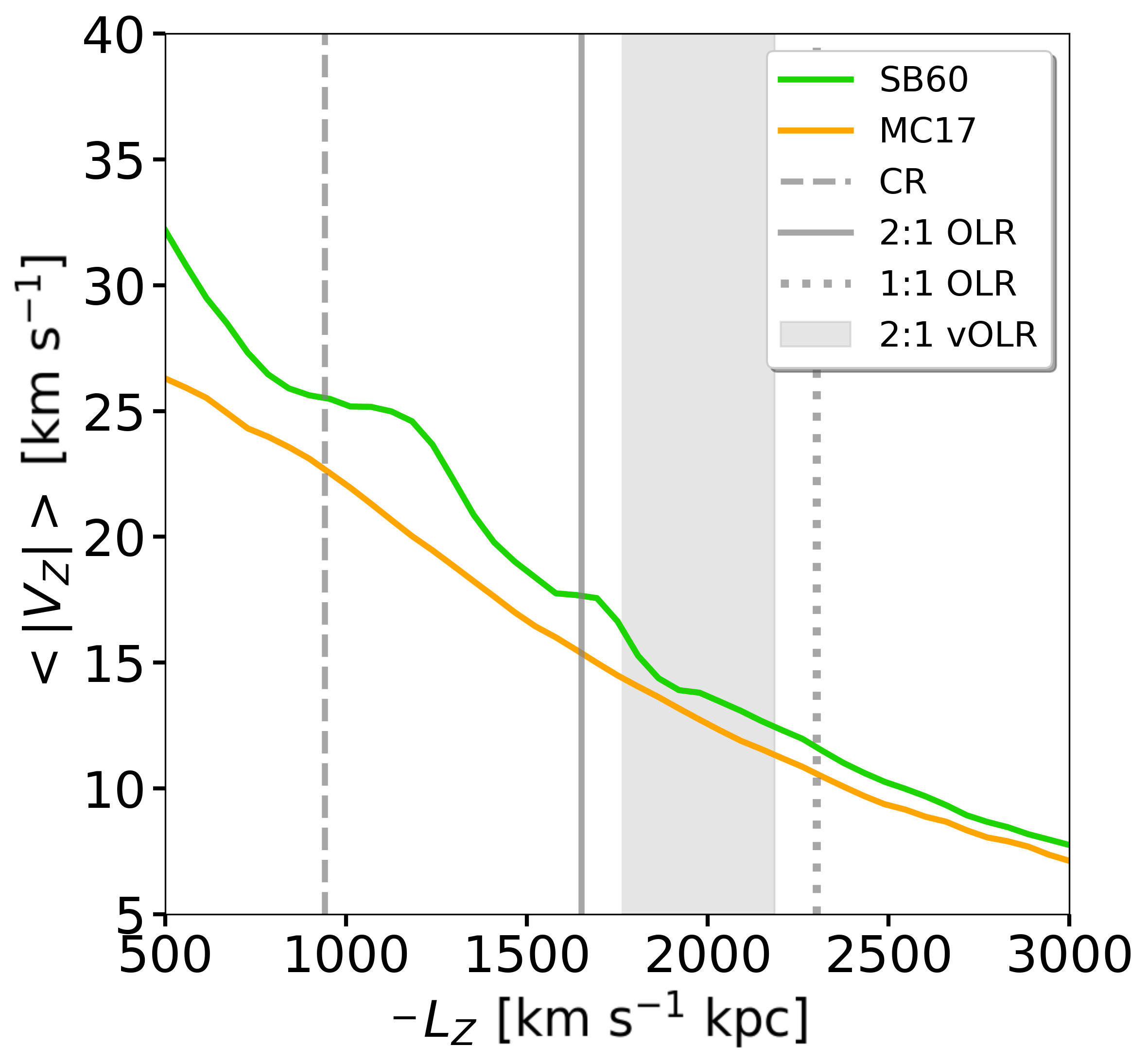}\\
\includegraphics[width=.38\textwidth]{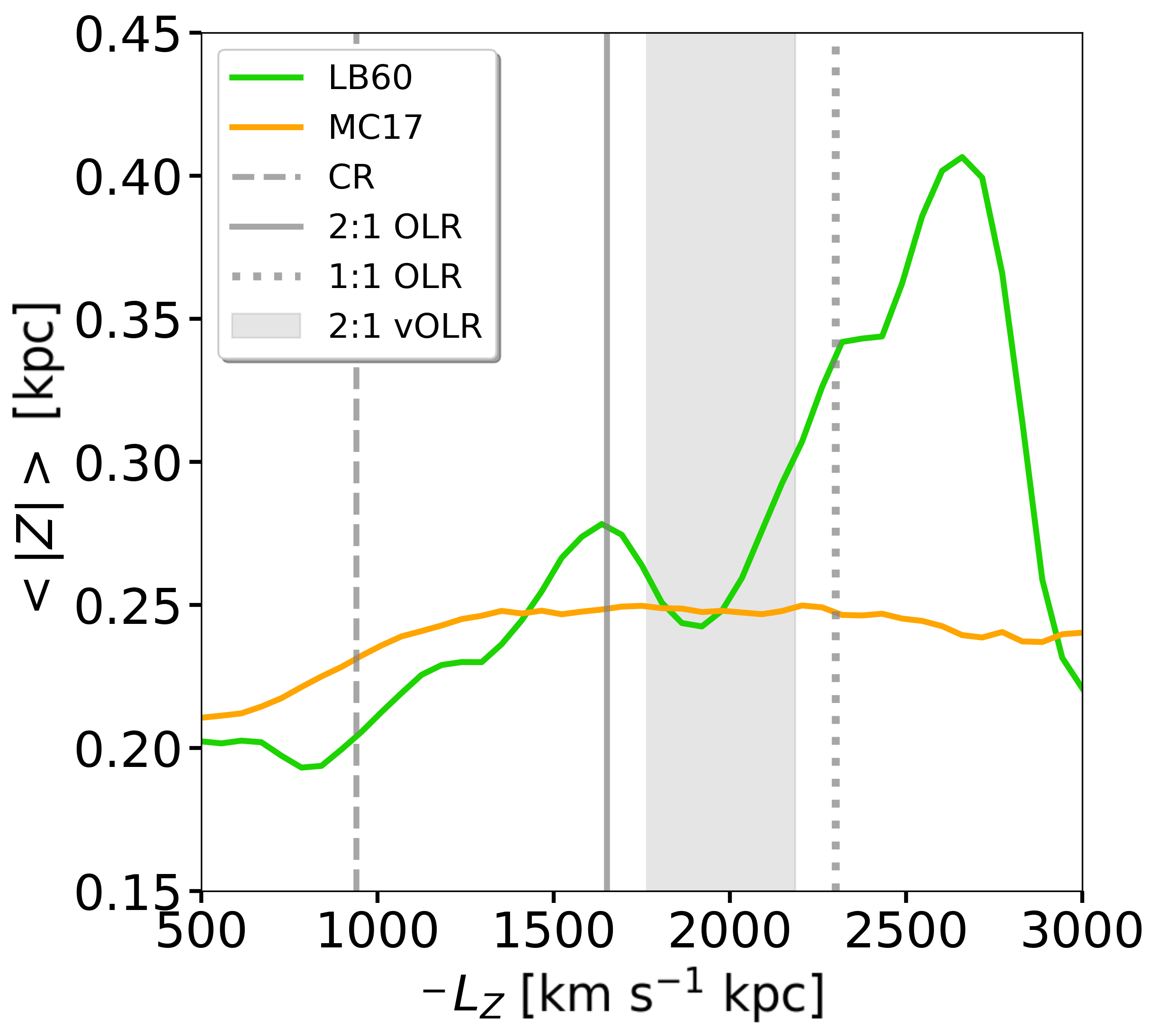}
\includegraphics[width=.37\textwidth]{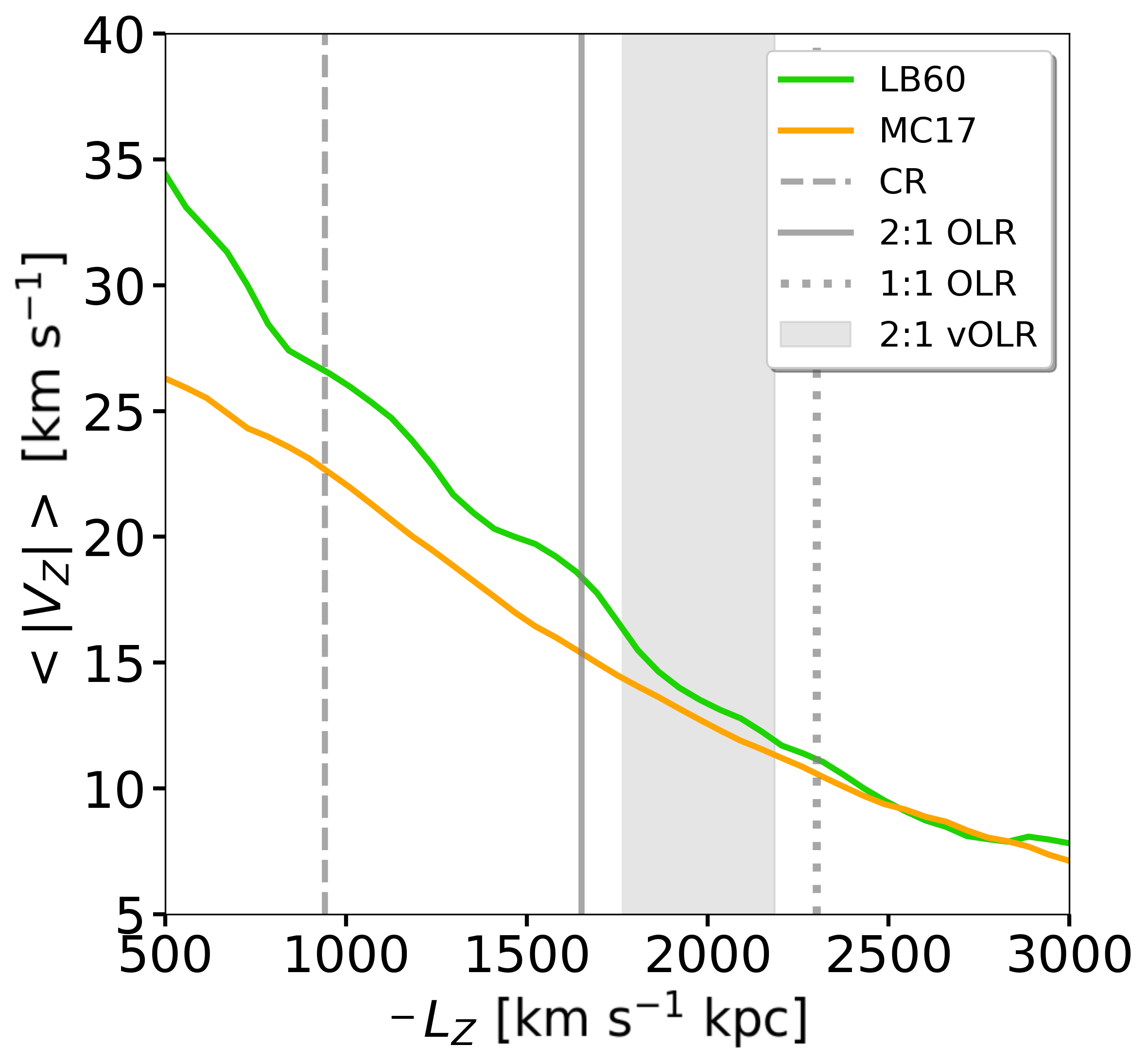}\\
\includegraphics[width=.38\textwidth]{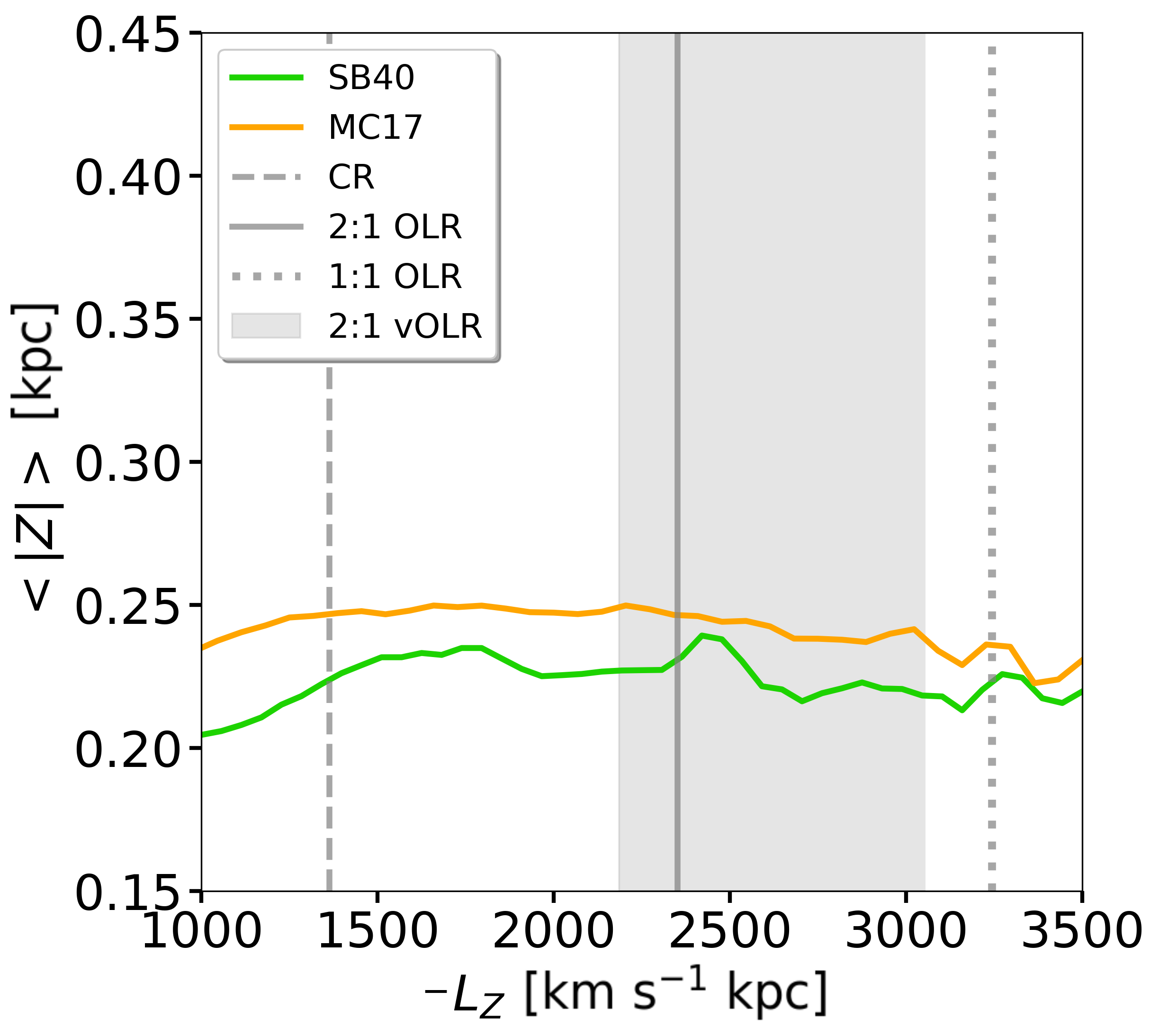}
\includegraphics[width=.37\textwidth]{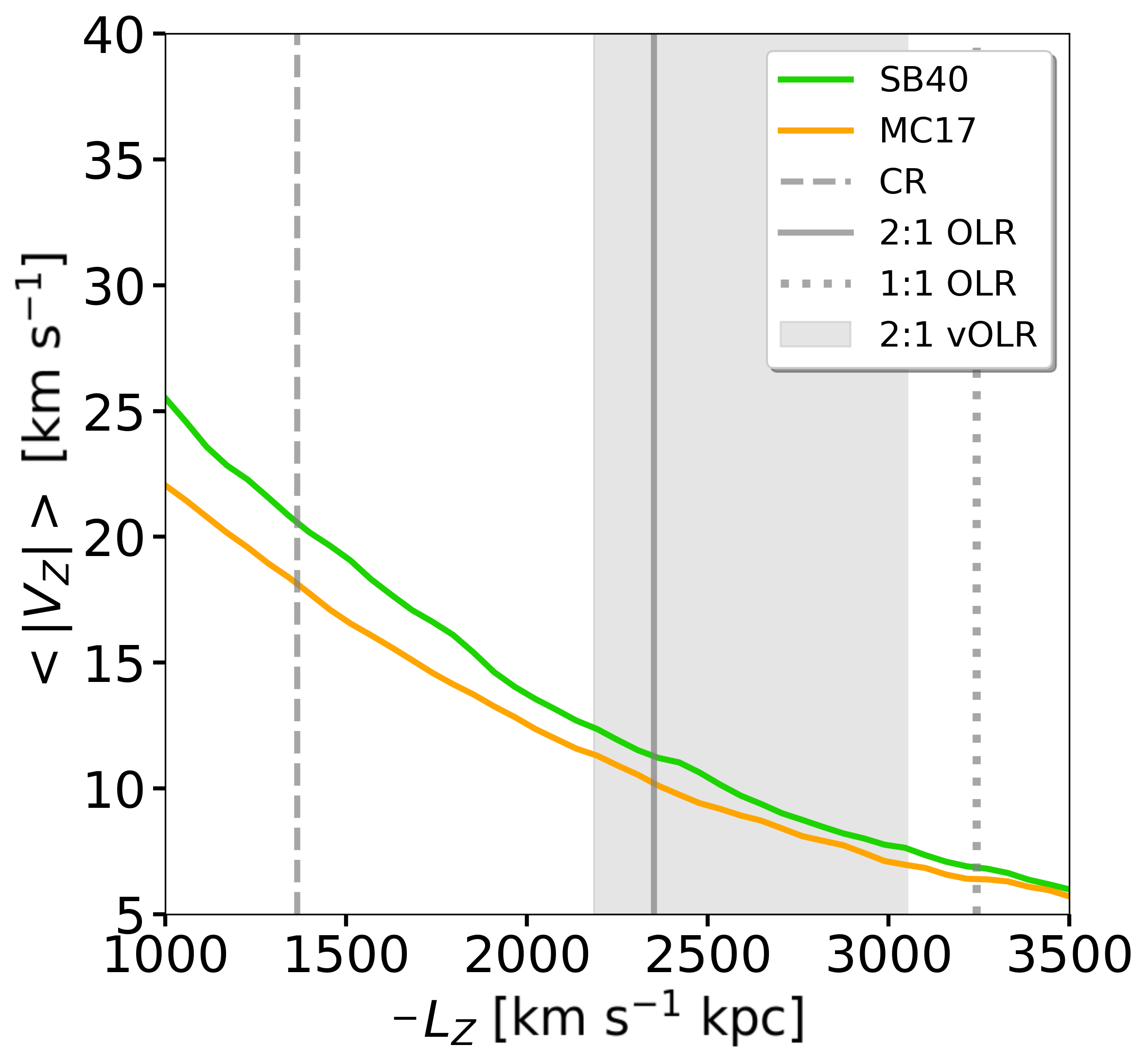}\\
\includegraphics[width=.38\textwidth]{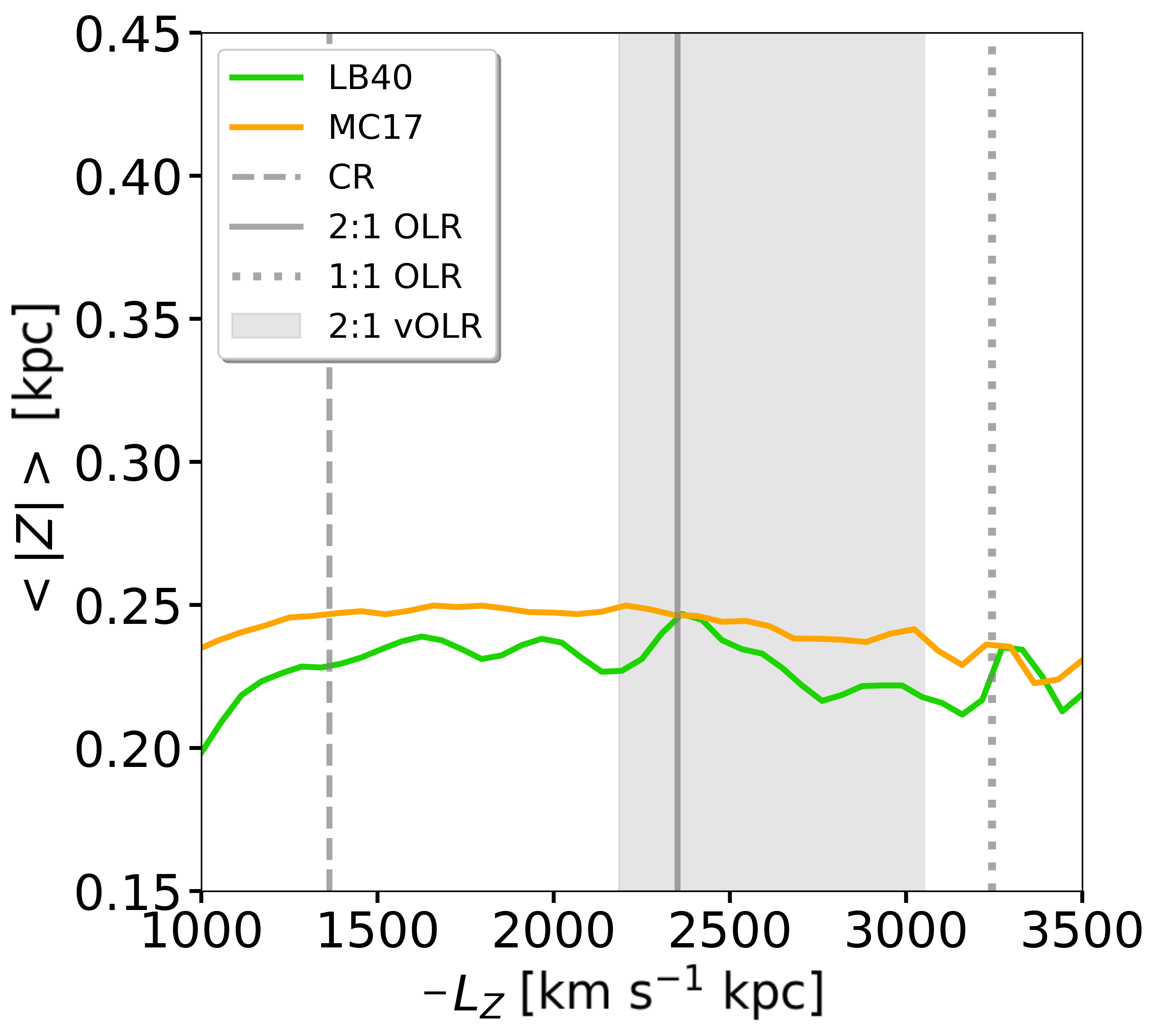}
\includegraphics[width=.37\textwidth]{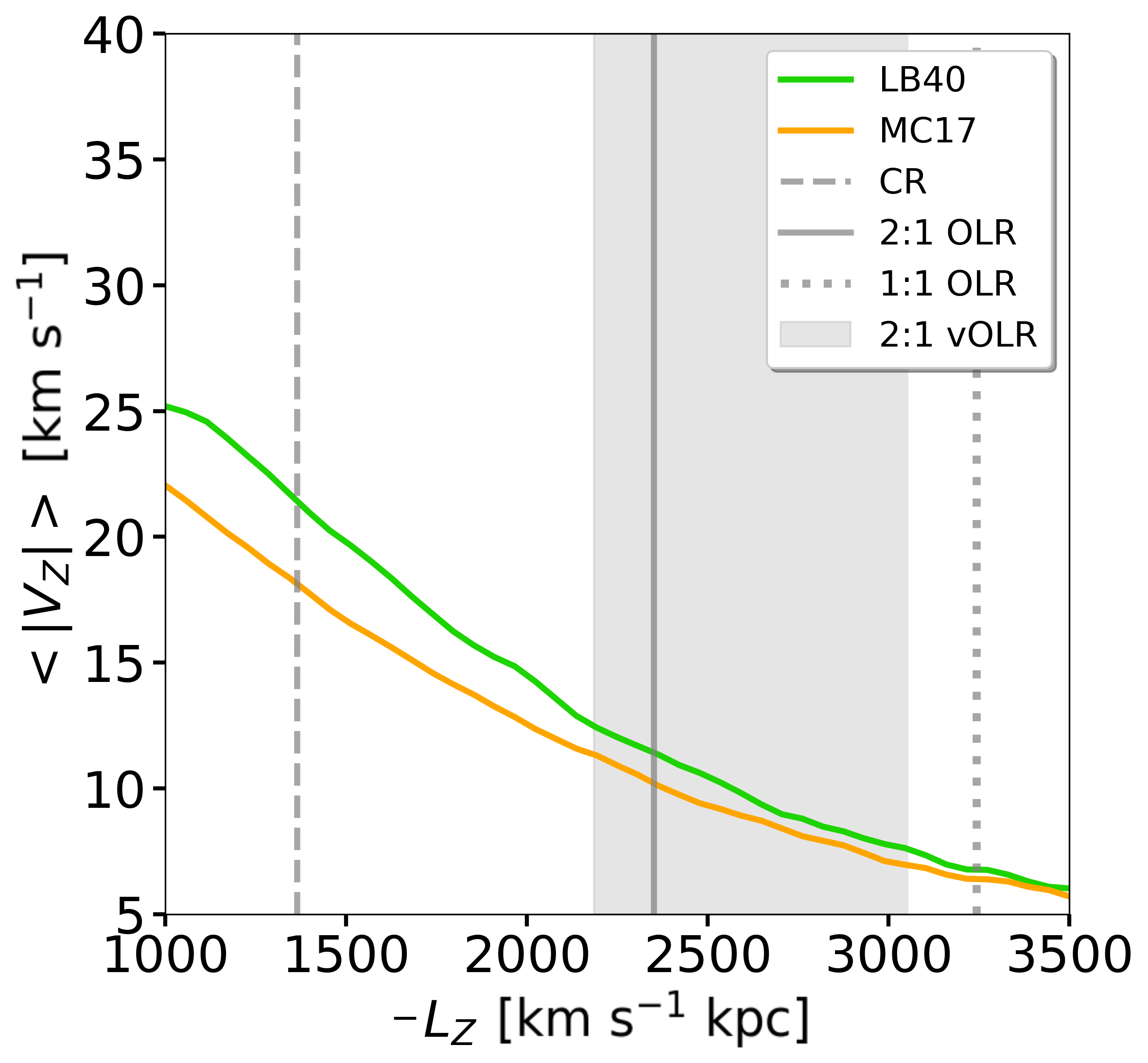}
\caption{Green lines---average coordinate $<|Z|>$ (left column) and average velocity component $<|V_Z|>$ (right column) as functions of the angular momentum $L_Z$ after three Gyr of evolution in barred potentials (SB60-first row, LB60-second row, SB40-third row and LB40 fourth row). {Orange lines are the same as for green ones, but for the model MC17.}
\label{figLZvert}} 
\end{figure}

\begin{figure}[H]
\begin{adjustwidth}{-\extralength}{0cm}
\centering
\includegraphics[width=4.4cm]{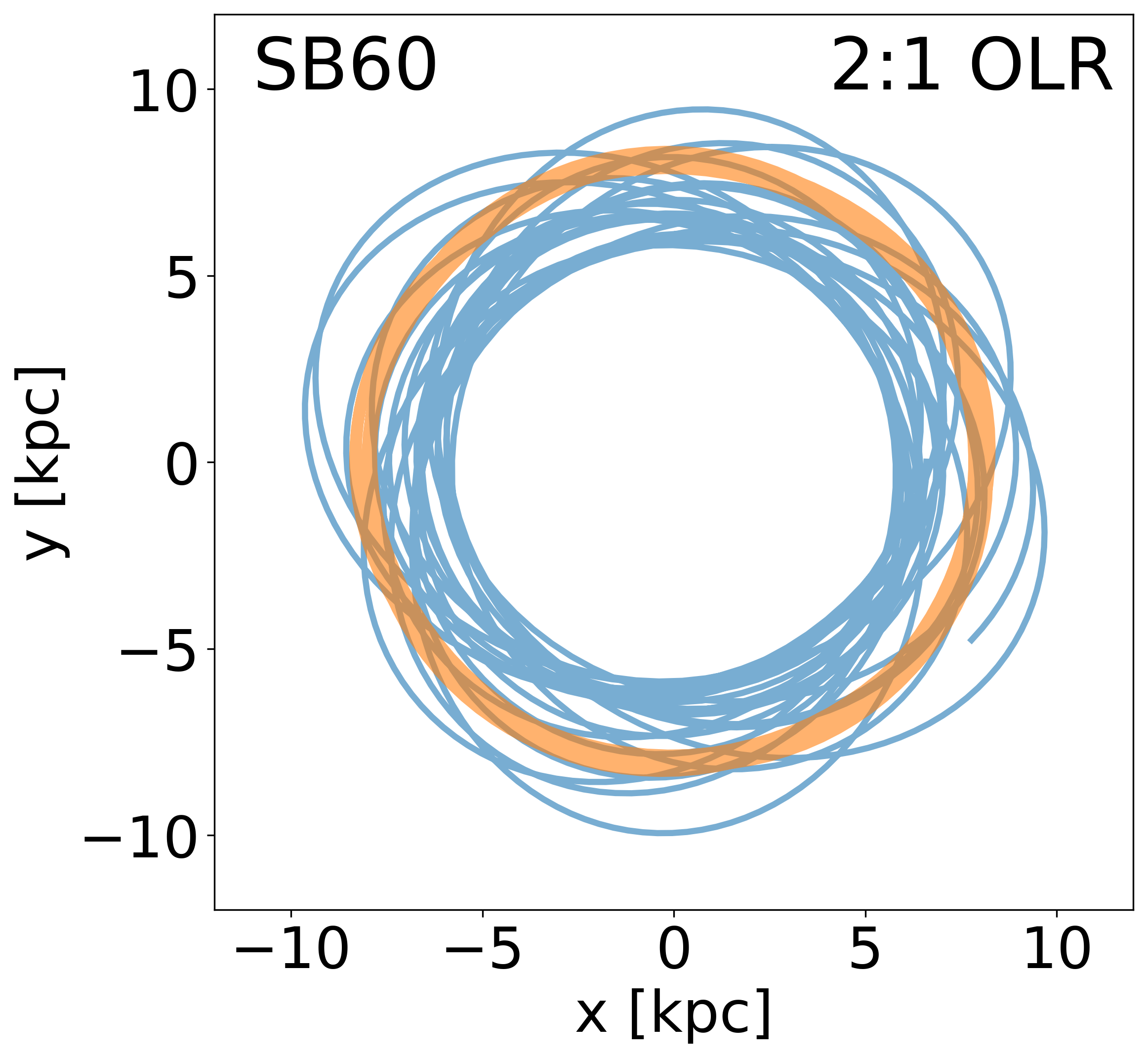}
\includegraphics[width=4.4cm]{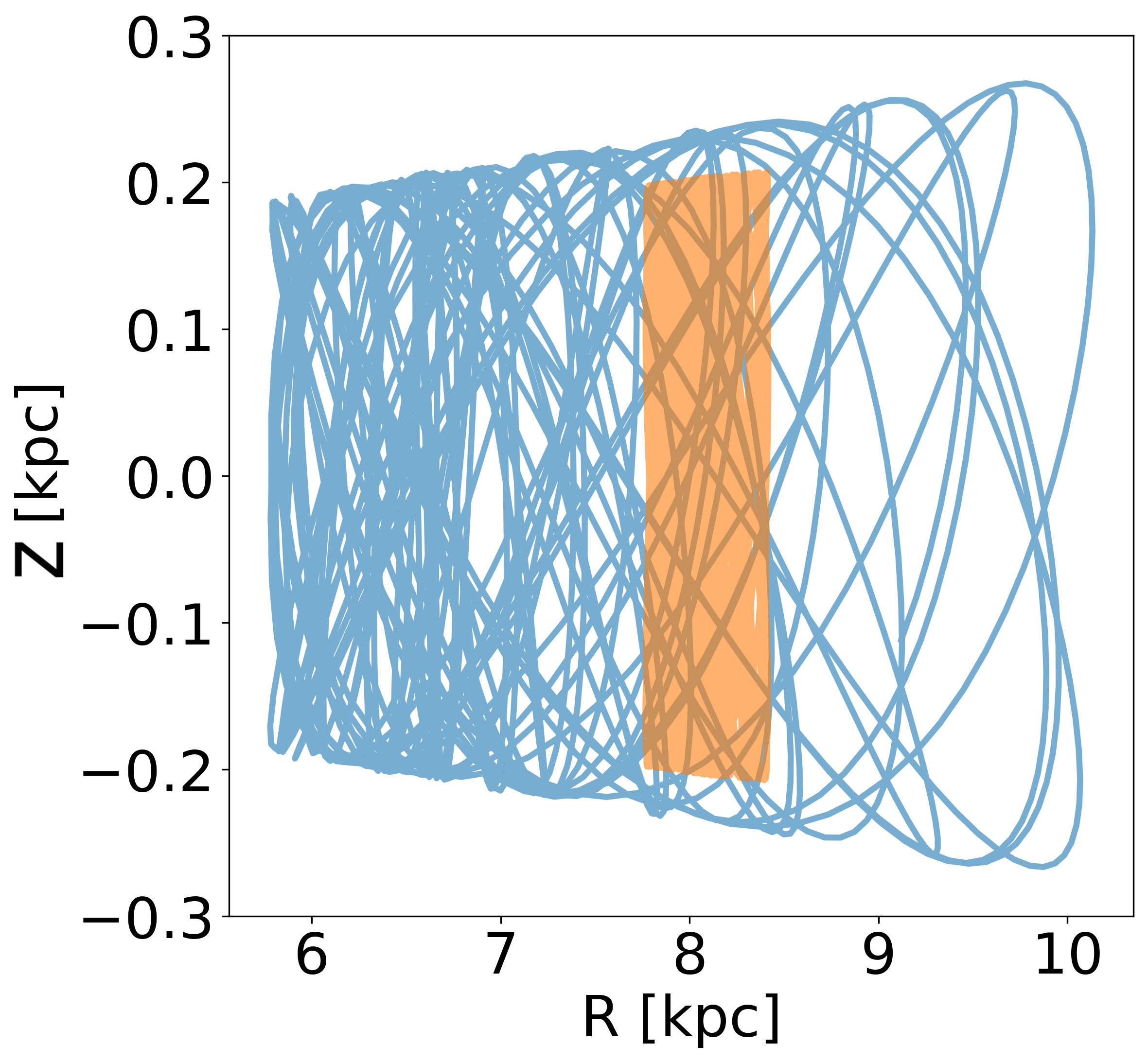}
\includegraphics[width=4.4cm]{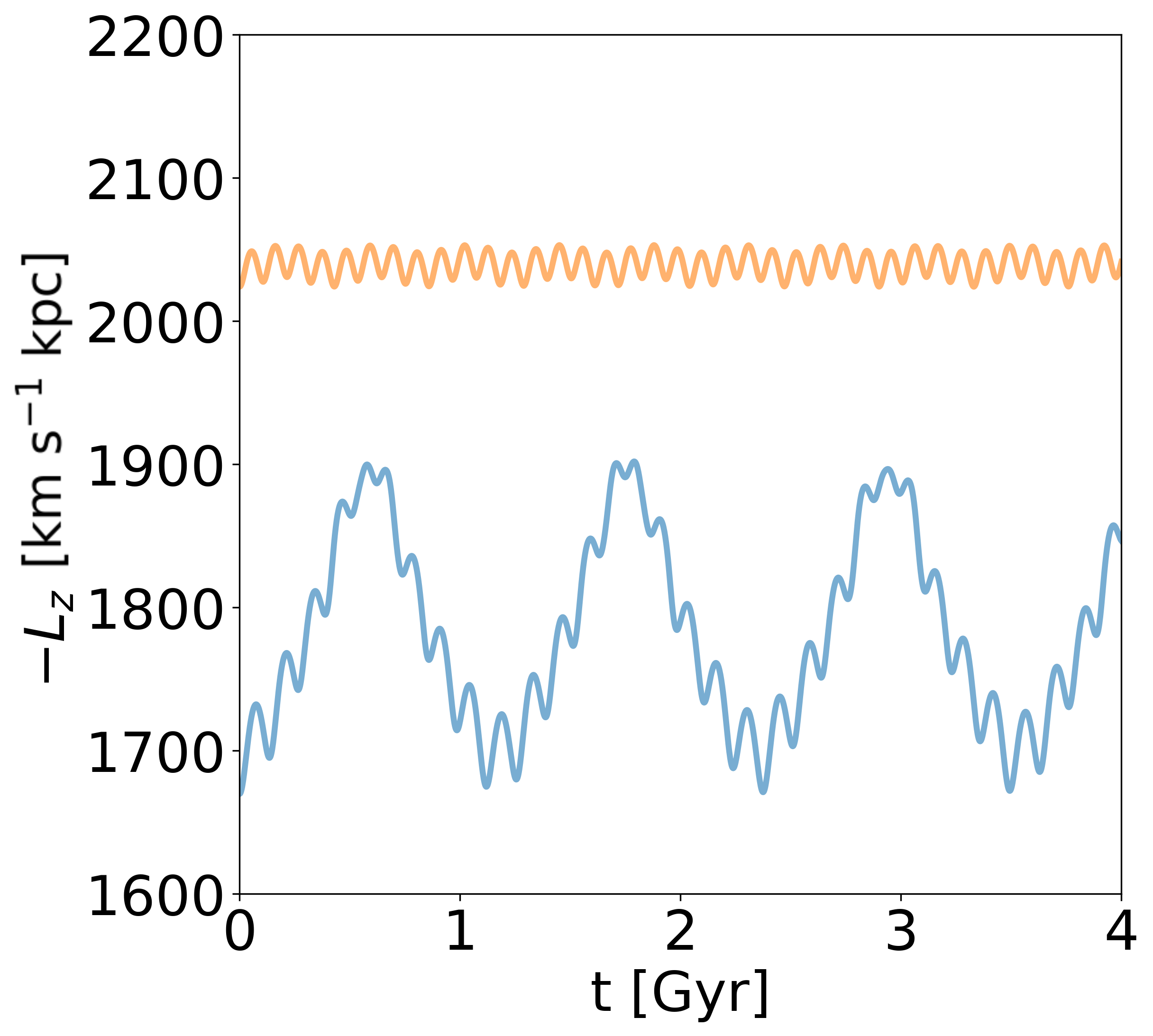}
\includegraphics[width=4.2cm]{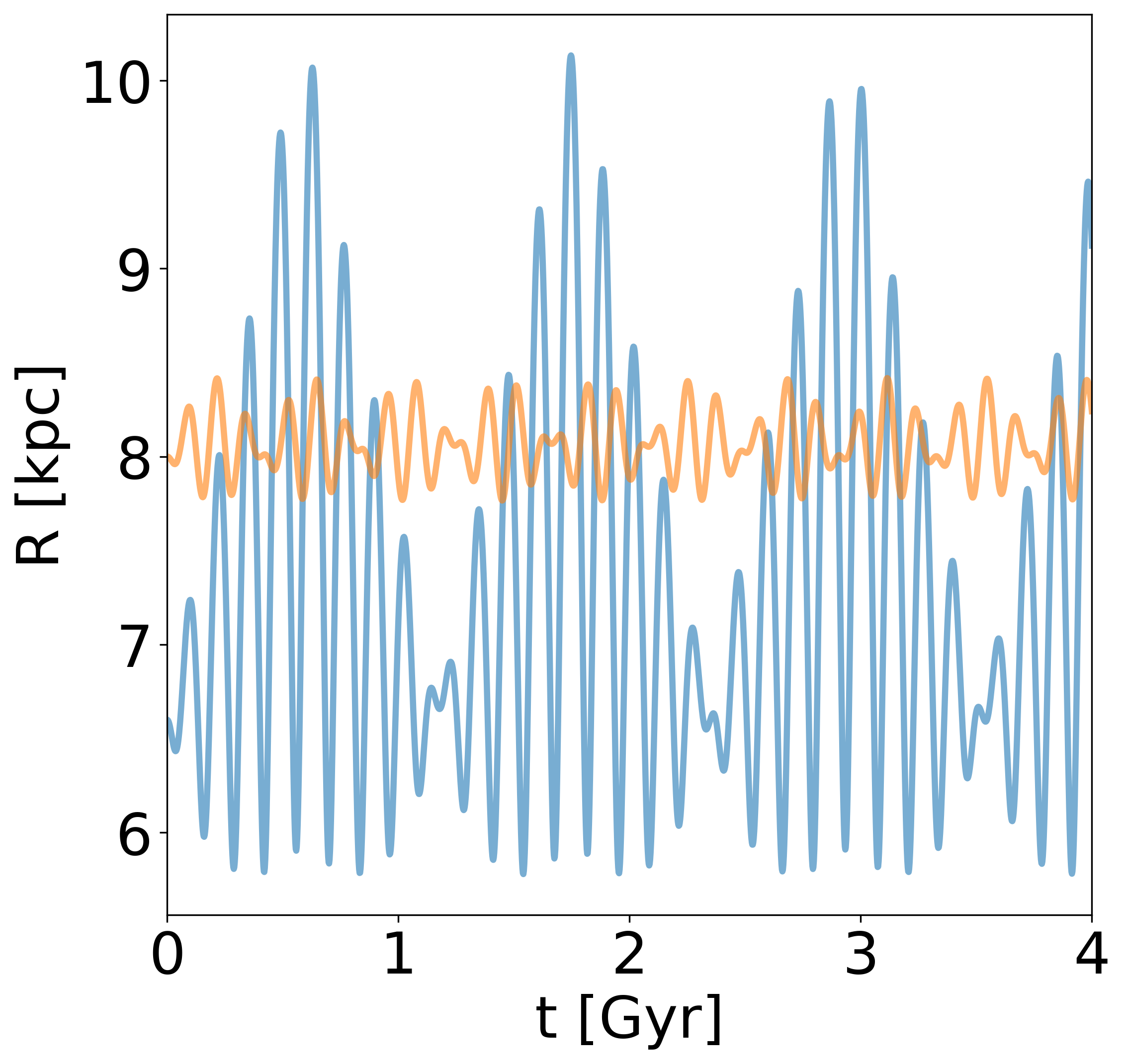}
\includegraphics[width=4.4cm]{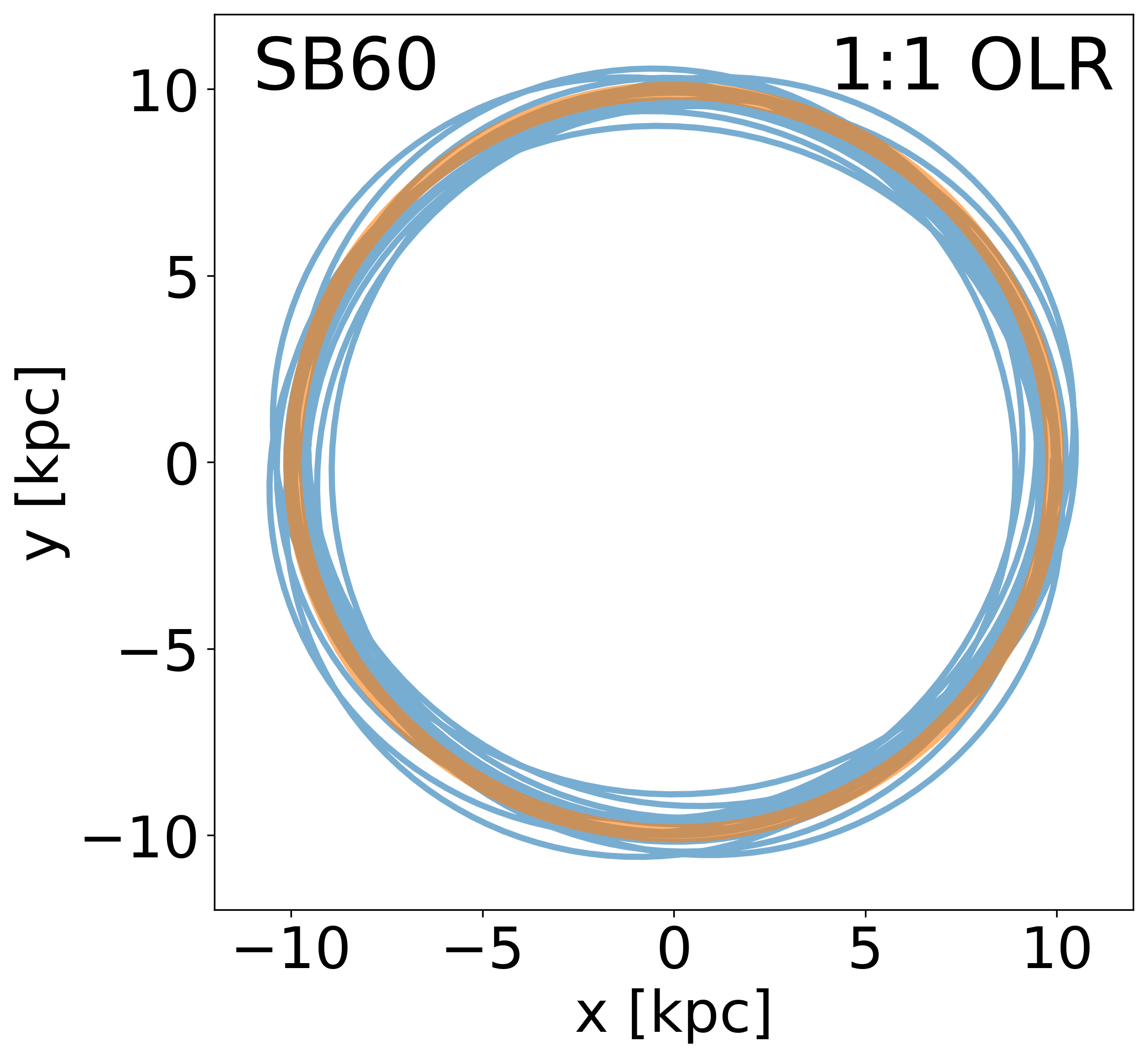}
\includegraphics[width=4.4cm]{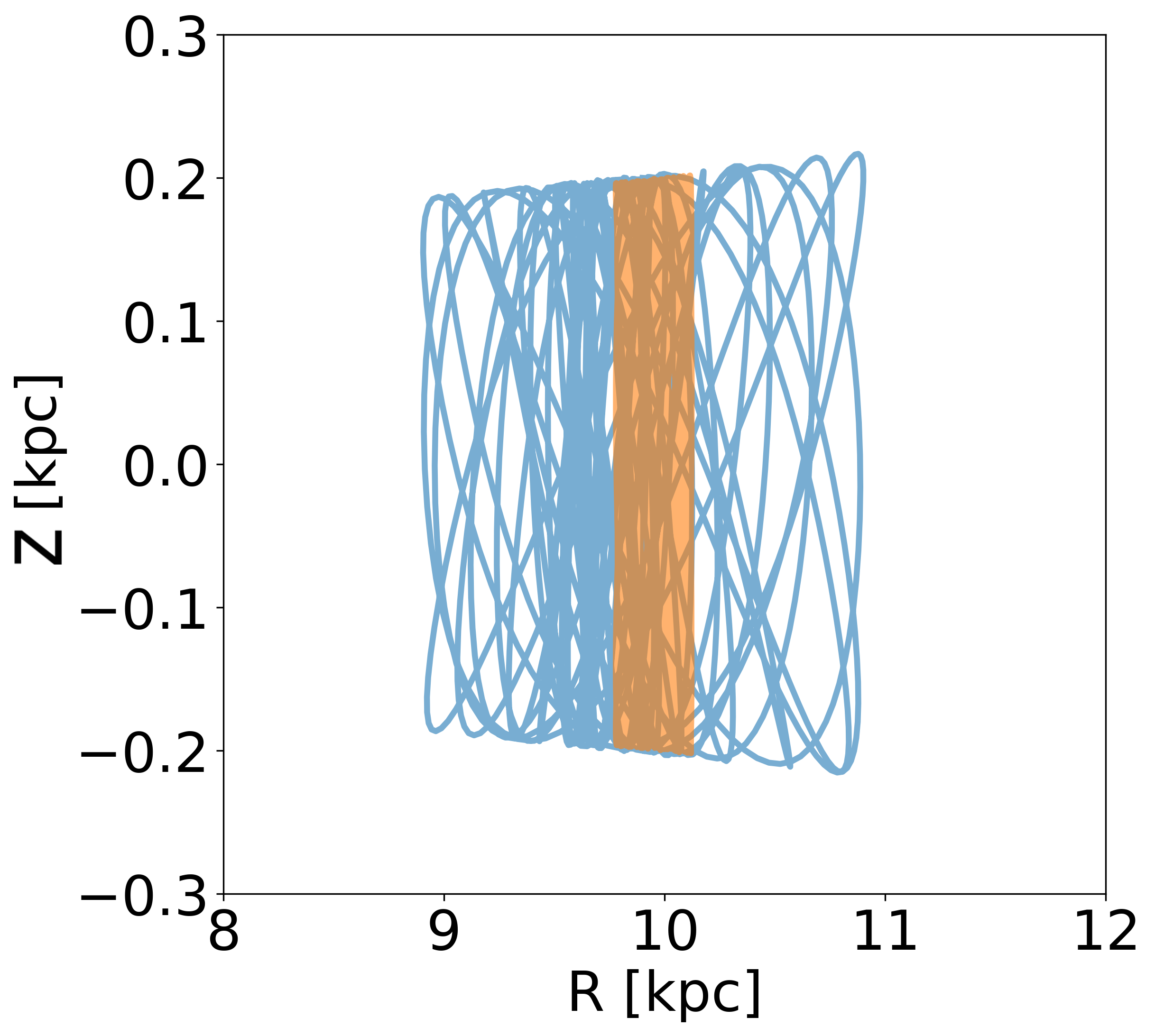}
\includegraphics[width=4.4cm]{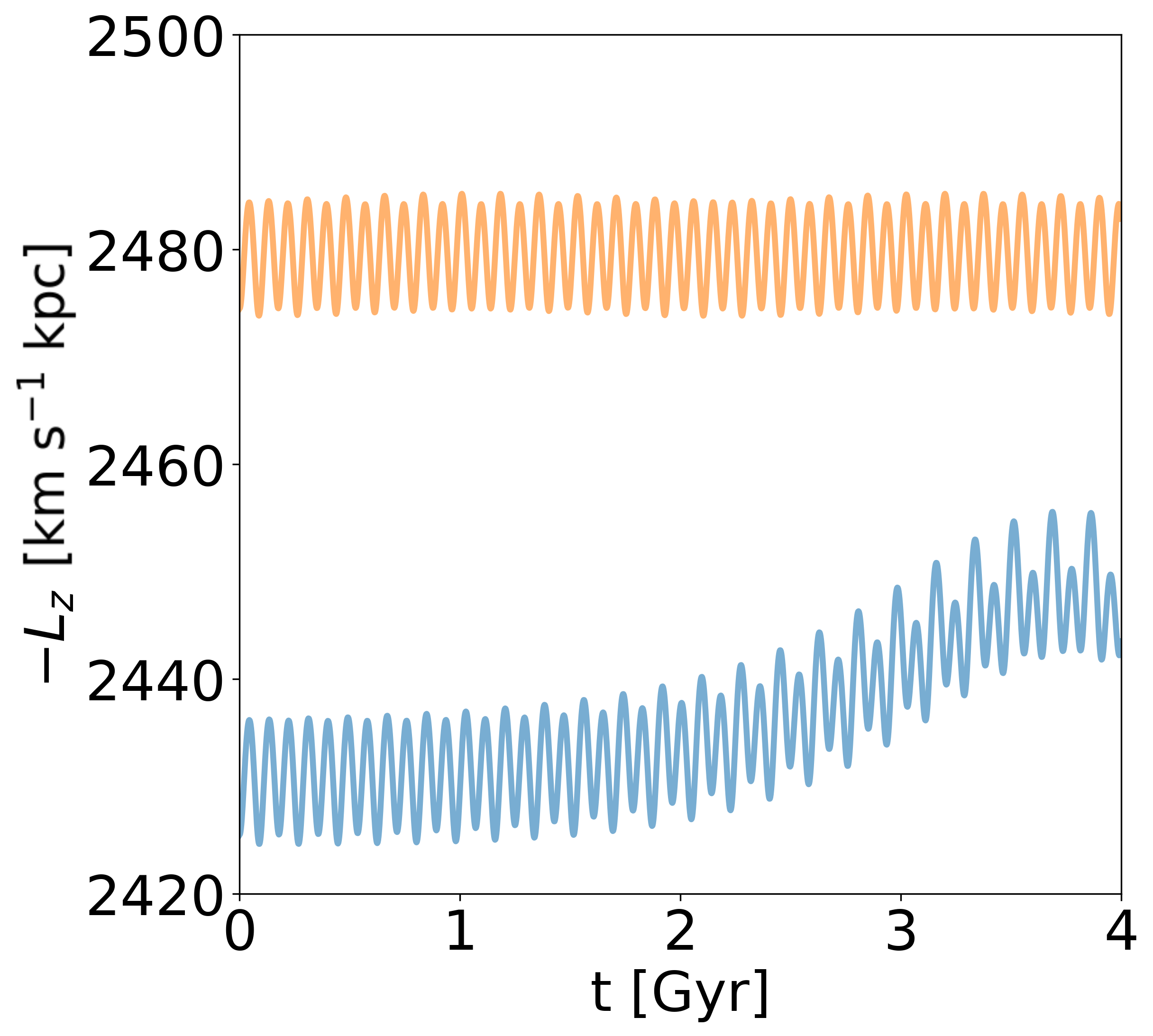}
\includegraphics[width=4.2cm]{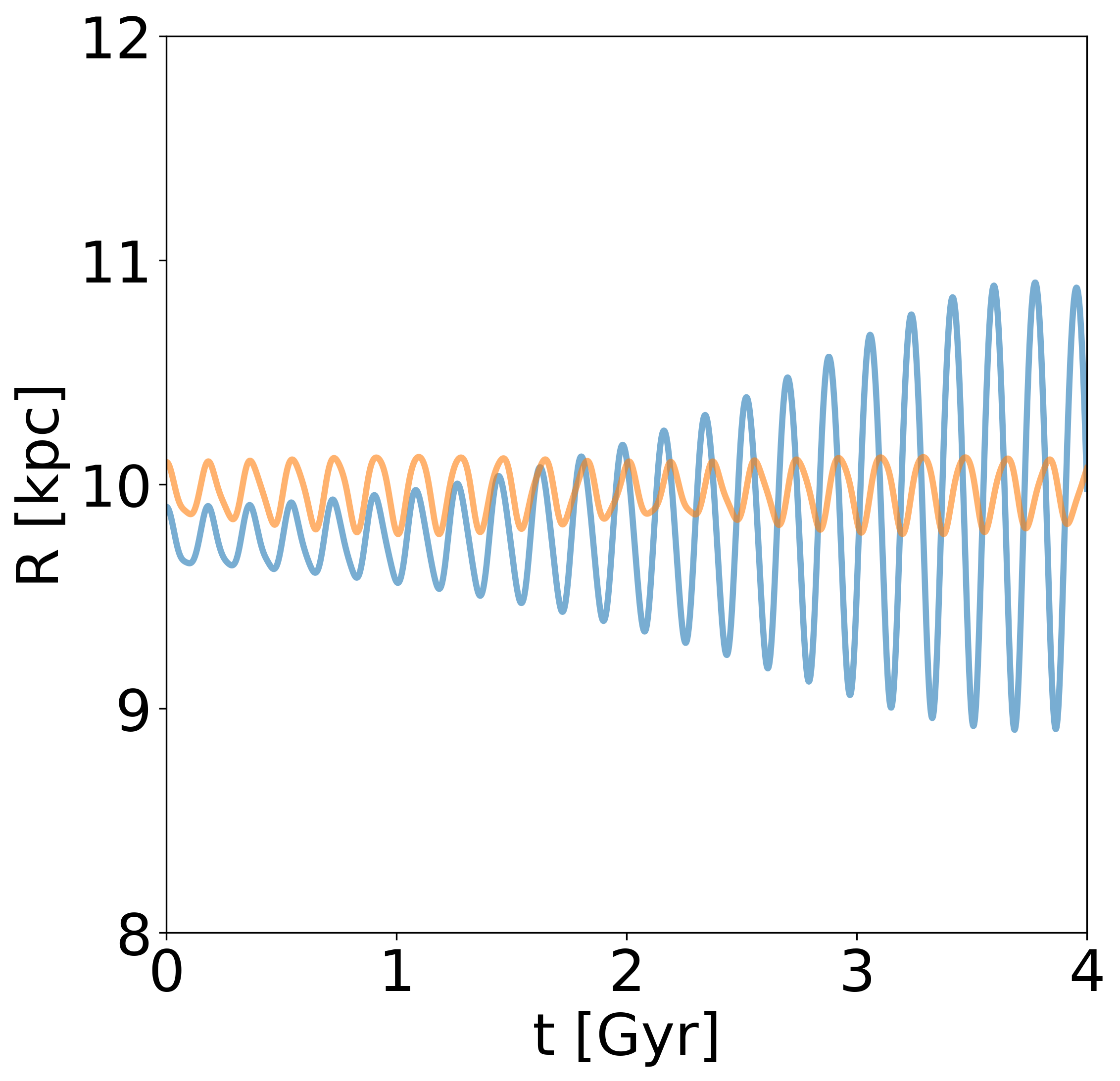}
\end{adjustwidth}
\caption{Evolution of resonant (blue) and non-resonant (orange) particles in model SB60 shown on $(x,y)$, $(R,Z)$, $(t,-L_Z)$ and $(t,R)$---planes at 2:1 (top row) and 1:1 (bottom row) OLR. 
\label{OLRS}} 
\end{figure}

To demonstrate the influence of resonances occurring in the plane of the disk on the disk's vertical structure, we follow
\citet{Trick} and describe the dynamics of the disk in $(L_Z,<|V_Z|>)$ and $(L_Z,<|Z|>)$ space.
Figure~\ref{figLZvert} shows $(L_Z,<|Z|>)$ and $(L_Z,<|V_Z|>)$ distributions after three Gyr of evolution of the models for moving averages of the values  $|Z|$ and $|L_Z|$.
As is seen from Figure~\ref{figLZvert}, the position of 2:1 OLR resonance correlates with the bumps on the $|Z|$ distribution of particles in all models. The dependence of the velocity component $V_Z$ on radius is also affected by bar at the 2:1 OLR resonance region. {The upper right panel of Figure~\ref{figLZvert} shows an elevation of $<|V_Z|>$ velocity in the vicinity of 2:1 outer Lindblad resonance in the model with the fast rotating long bar.}
A similar result was reported by \citet{Trick} (see their Figure~\ref{OLRS}c), who also demonstrated that the 2:1 OLR resonance with the bar can contribute to the peculiarities in distribution of $V_Z$ along the $L_Z$. To find the location of the resonances in $L_Z$ space in Figure~\ref{figLZvert}, we use the radial positions corresponding to the axisymmetric potential and the velocities corresponding to the rotation curve of MC17.

In models with a slowly rotating bar (SB40 and LB40), the influence of the 2:1 OLR on disk dynamics in the perpendicular direction is less 
noticeable due to its further location from the galactic center. The $Z$-distribution of the particles along the radius is perturbed by a few percent around the 2:1 outer Lindblad resonance. 
The influence of the 2:1 OLR on the particles $V_Z$ distribution alone is negligible.

To clarify in more detail the influence of resonances on $(-L_Z,<|Z|>)$ and \linebreak  ($-L_Z,<|V_Z|>$) distributions, we trace the dynamics of the resonant and non-resonant particles in the model with the short fast rotating bar SB60. Figure~\ref{OLRS} shows the evolution of resonant (blue) and non-resonant (orange) particles at 2:1 (top row) and 1:1 (bottom row) outer Lindblad resonances.
As one can see from the  $L_Z - t $ plot in Figure~\ref{OLRS}, the angular momentum of resonant particles shows an oscillational behavior atop of the smaller time-scale oscillations, corresponding to the periodic disturbances from the bar, i.e., there is an oscillational behavior with a larger period. Similar behavior of the particles at the 2:1 outer Lindblad resonance was reported by \citet{Melnik}, who gave an explanation for such behavior. Such behavior of the resonant particles influences the properties of the disk in the Z-direction.  As one can see from Figure \ref {OLRS}, the resonant particles can deviate up to 10 kpc from the resonance region, located at $\approx$ 6.5 kpc, bringing the particles with 
higher velocity dispersion $\sigma_z$ to the regions with smaller values of z-component of the velocity dispersion of the disk, which leads to the appearance of a bump on $<Z>$--$L_Z$ distribution in Figure~\ref{figLZvert}.

\begin{figure}[H]
\includegraphics[width=.42\textwidth]{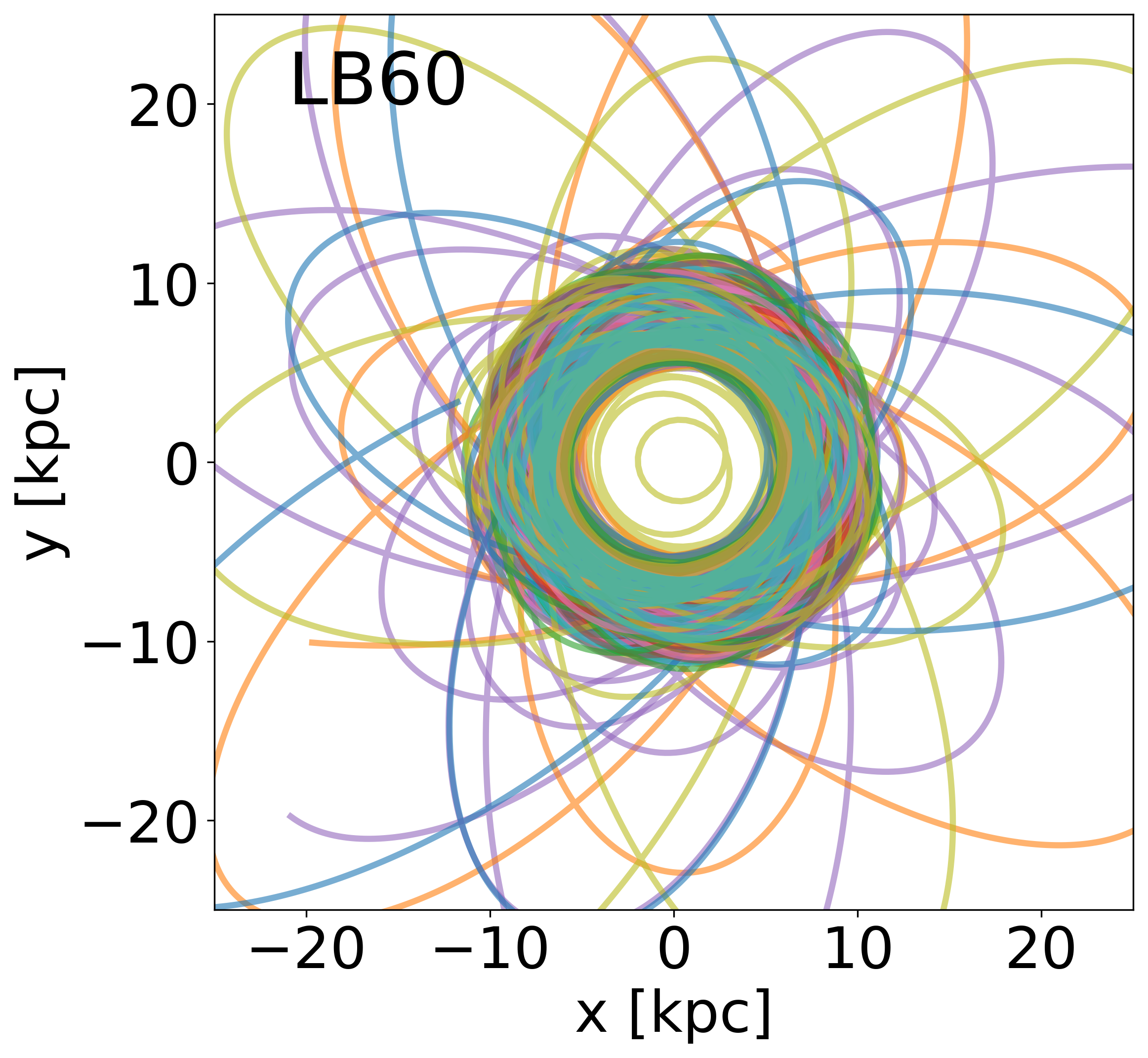}
\includegraphics[width=.44\textwidth]{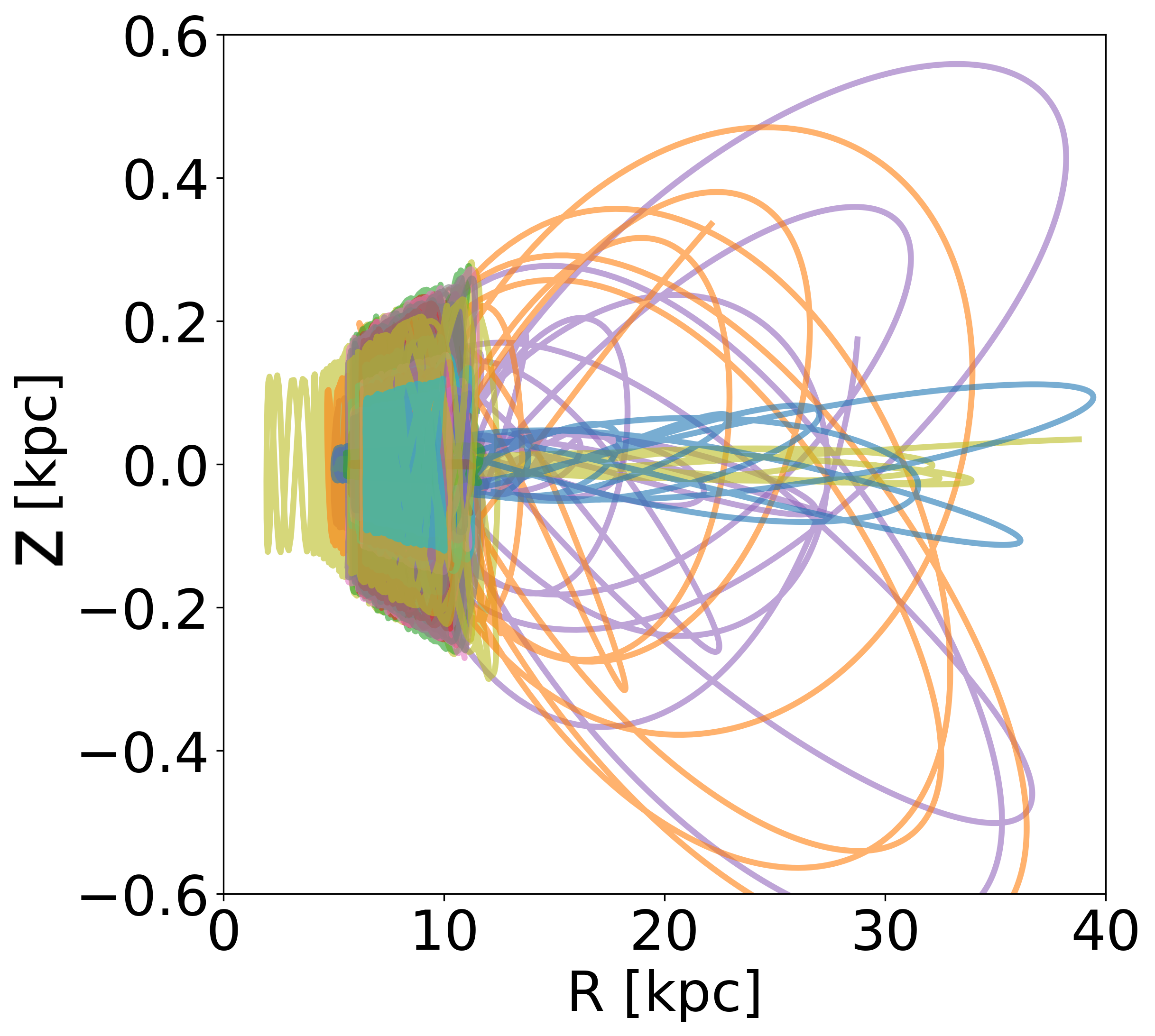}\\
\includegraphics[width=.44\textwidth]{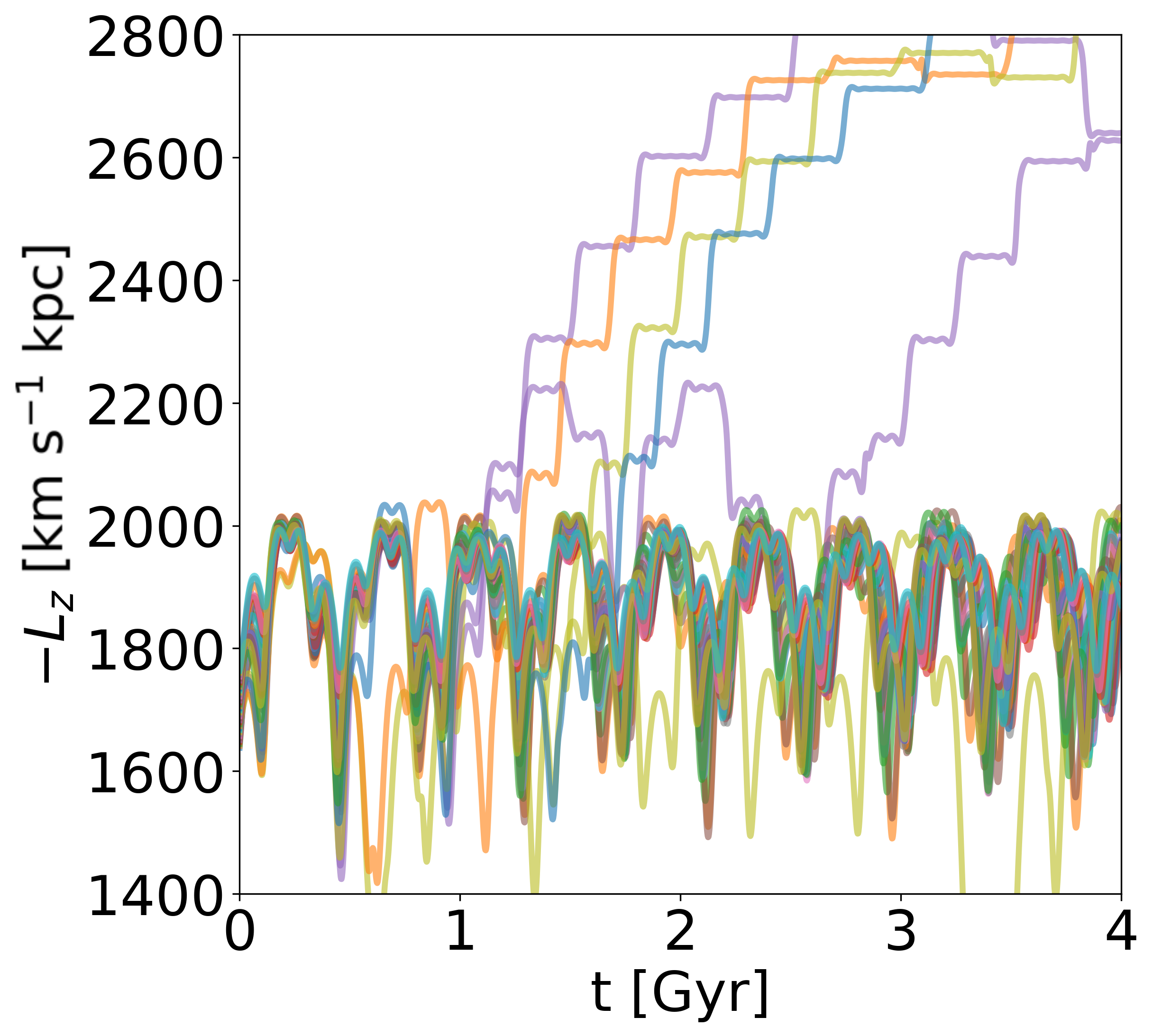}
\includegraphics[width=.42\textwidth]{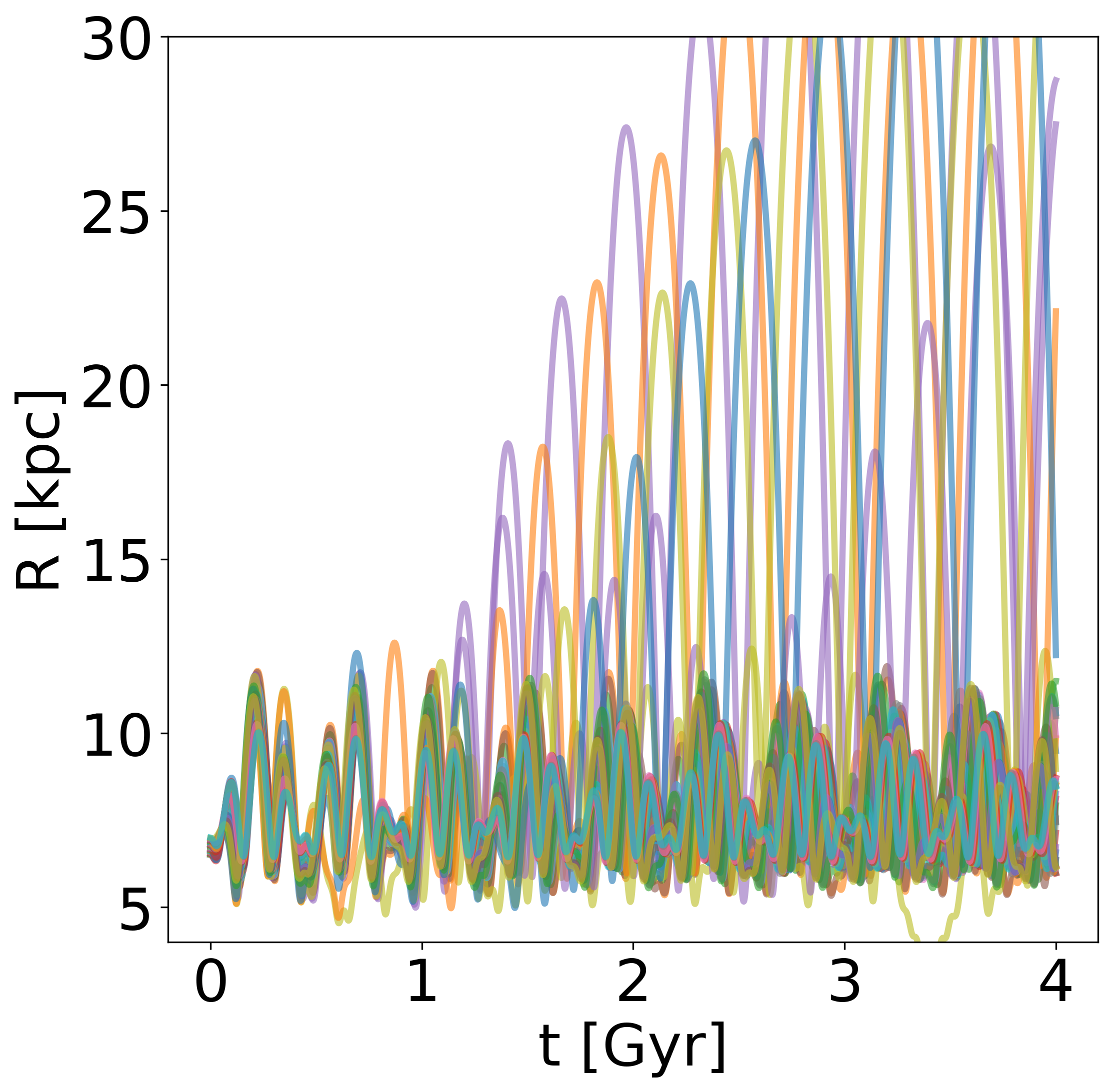}
\caption{Orbit evolution of escaping particles in the model LB60 shown on $(x,y)$, $(R,Z)$, $(t,-L_Z)$ and $(t,R)$-planes.  
\label{Escapers}} 
\end{figure}

Let us discuss in more detail  the  $(-L_Z,<|Z|>)$-distribution of the particles in LB60 model.  Figure~\ref{figLZvert} shows that  particles with angular momenta of $\approx$  2200--2800 $~\mbox{km}~\mbox{s}^{-1}~\mbox{kpc}$ considerably increase the thickness of the disk with an average growth of the Z-coordinate of one hundred percent. 
The appearance of a bump on $(-L_Z,<|Z|>)$-diagram is related to the 
'escapers'---the particles at the 2:1 OLR region that have amplitudes of oscillations large enough so the particles reach the bar region, and get an additional ``kick'' from a bar.  Figure~\ref{Escapers} supports this picture. As one can see from $(t,L_Z)$ diagram of this figure, most of the resonant particles have a 'typical' behavior at the resonant region with angular momenta nonlinearly oscillating around the equilibrium value.  A few percent of the particles that closely approach the bar region  increase
their angular momenta from from $-L_Z\approx$ 1600--1800 to 2200--3000 $~\mbox{km~s}^{-1}~\mbox{kpc}$, reaching as a consequence large distances, both in radius and in the direction perpendicular to the disk. \citet{escp} reported about the formation of the escaping particles from the OLR-region due to an interaction with a slowing bar. Our simulations show that the formation of escapers occurs in a constantly rotating bar potential as well.

\begin{figure}[H]
\begin{adjustwidth}{-\extralength}{0cm}
\centering
\includegraphics[width=4.4cm]{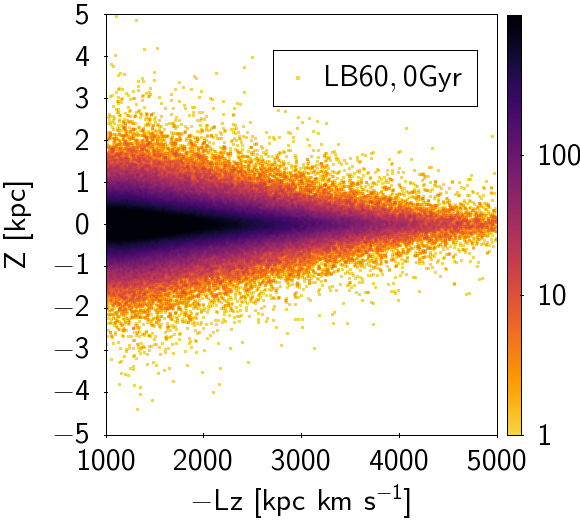}
\includegraphics[width=4.4cm]{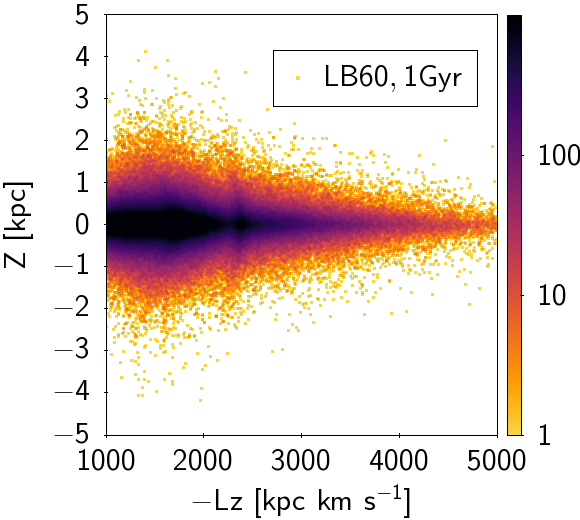}
\includegraphics[width=4.4cm]{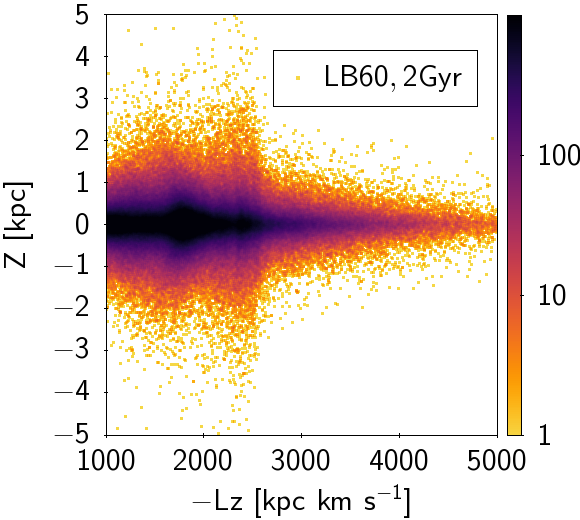}
\includegraphics[width=4.4cm]{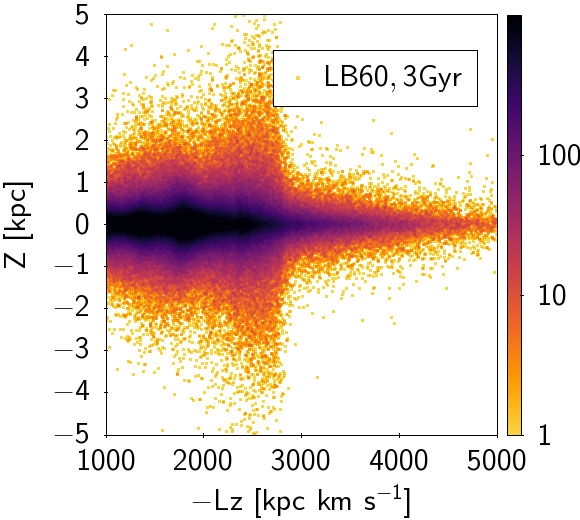}
\includegraphics[width=4.4cm]{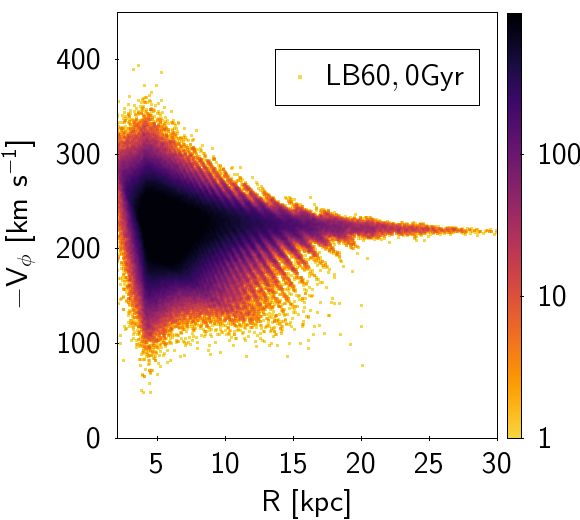}
\includegraphics[width=4.4cm]{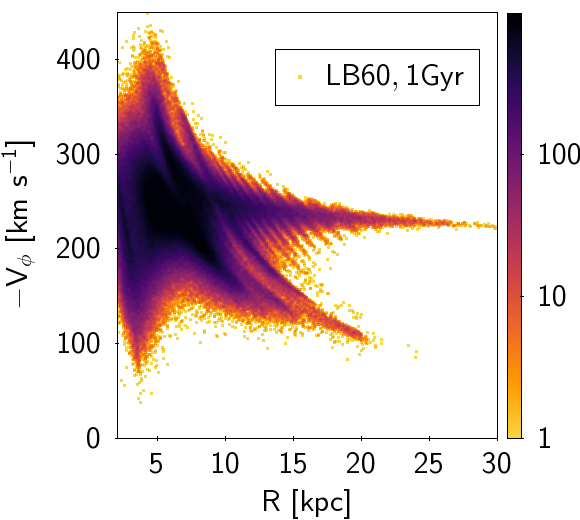}
\includegraphics[width=4.4cm]{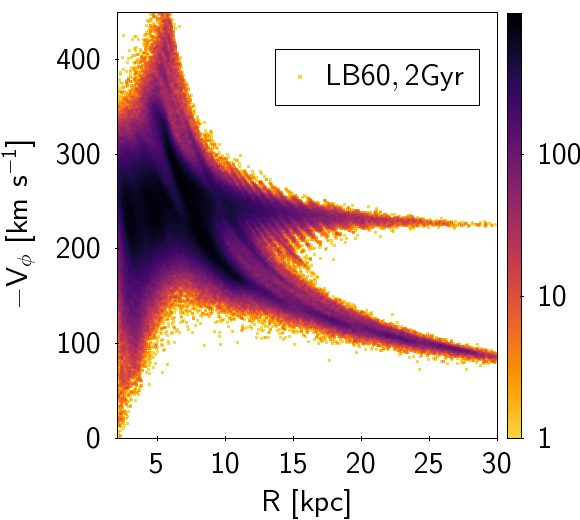}
\includegraphics[width=4.4cm]{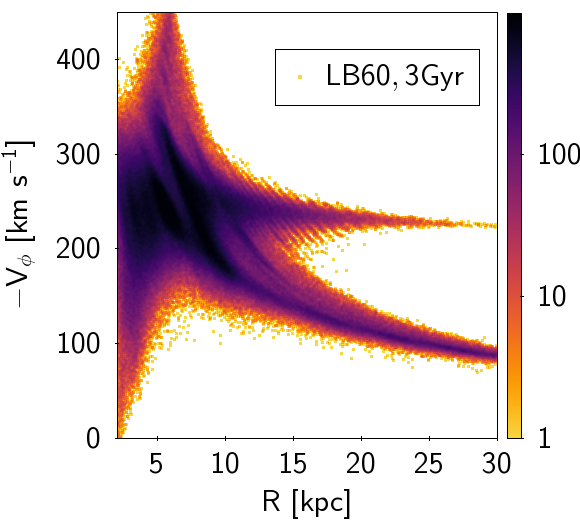}
\includegraphics[width=4.4cm]{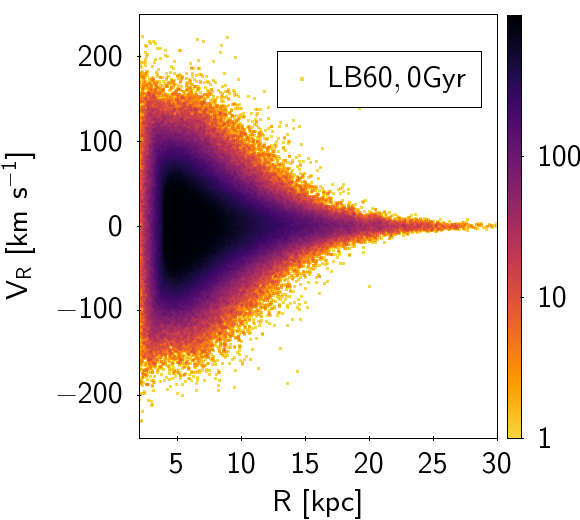}
\includegraphics[width=4.4cm]{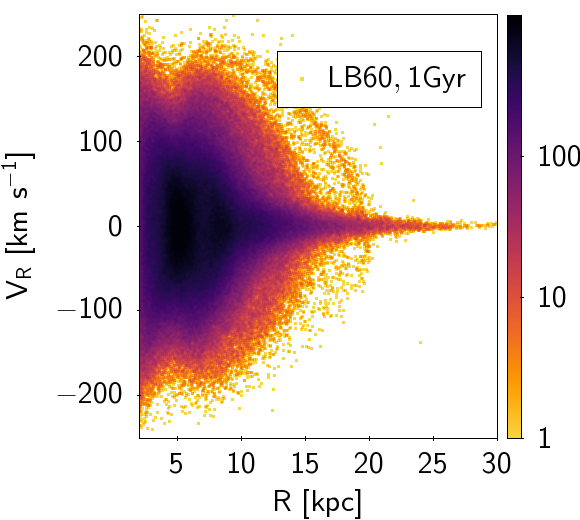}
\includegraphics[width=4.4cm]{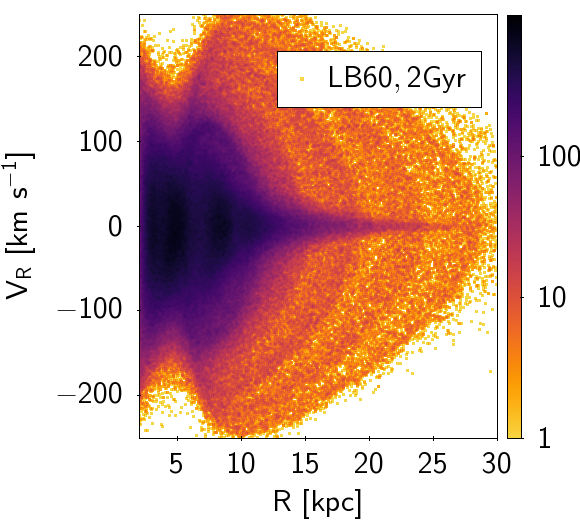}
\includegraphics[width=4.4cm]{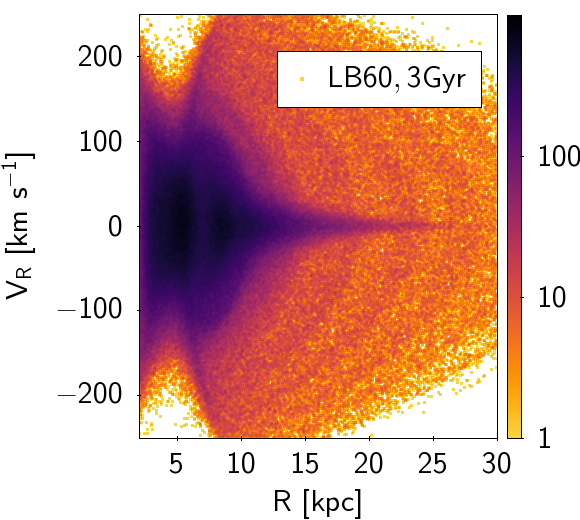}
\end{adjustwidth}
\caption{Evolution of $(Z)$ (first row), $(V_{\phi})$ (second row), and $V_{R})$ (third row) distributions of particles in LB60 model.\label{figVerRes}}
\end{figure}

The dynamics of resonant particles in the model LB60 with  the fast elongated bar has some other features not seen in the models
with a shorter (SB60) or a slowly rotating bar (SB40 and LB40). 
Figure~\ref{figVerRes} shows evolution of the particles in the LB60 model shown in 
$(L_Z,Z)$, $(R,V_{\phi})$ and $(R,V_{R})$-planes over three Gyr of evolution.  
The $(L_Z,Z)$-plane of this figure demonstrates growth of the resonant structure for  $L_Z\approx$ 2000--3000 $~\mbox{km}~\mbox{s}^{-1}~\mbox{kpc}$. 
We already discussed this feature also seen in Figure~\ref{figLZvert}.

Another manifestation of escaping particles, shown in Figure~\ref{Escapers}, is the growth with time of high-velocity and low-velocity 
``tails'', seen on the $(R,V_{\phi})$ distributions of Figure~\ref{figVerRes}. The origin of the tails is related to the escaping particles, which,
due to the resonant set-up at the 2:1 
outer Lindblad resonance, come close enough to the bar and deviate far from their original locations. 
The orbits of such resonant particles, occupying a relatively narrow band of angular momenta  
$L_Z \approx \mbox{2200--2800}$$~\mbox{km}~\mbox{s}^{-1}~\mbox{kpc}$, become more elongated with time, with growing apocentric distances and pericentric velocities 
of the particles seen on the $(R,V_{\phi})$ diagram of Figure~\ref{figVerRes} as high- and low-velocity tails.

The bottom row of Figure~\ref{figVerRes} shows the formation of chevron-like structures in the $(R,V_{R})$ distribution of the particles interacting with the long fast bar. Chevron-like structures are observed in the velocity distribution of the Milky Way halo stars~\citep{Davies,Belokur}, and were recently discovered in the Andromeda galaxy by \citet{Dey}.  
\citet{Davies}, \citet{Belokur} and Wenbo Wu \cite{WenboWu} 
suggest that such structures can result from  accretion of a few satellite galaxies in the past of the Milky Way's history.
Interaction of stars with a bar at corotation resonance can also lead to the formation of chevron-like structures, as shown recently by \citet{Dillamore1}.
Our simulations demonstrate that chevrons can also be formed by the particles escaping from the 2:1 outer Lindblad resonance due to a strong interaction with the fast long bar. We find that about five percent of the resonant stars are involved in the escaping process, and the formation of the chevron-like structures and tails. The formation of chevron-like structures by the escapees from the OLR region, requires, however, a more detailed study.  

It should be noted that such effects appear only in the LB60 model, for which a fast rotation provides a 2:1 OLR resonance position closer to the centre (near 7 kpc), unlike the slow bar (10 kpc). The parameters of the Milky Way bar, such as the length of the semi-major axis and the angular velocity of rotation, are still largely uncertain and it is not possible to draw an unambiguous conclusion about whether the LB60 model is acceptable for our galaxy. But it should also be noted that in the observational data of other galaxies there are both very rapidly rotating bars and long bars~\citep{LeeB}, in which such resonant effects may take place. However, the observation of structures such as chevrons, or high and low-velocity ``tails'' remain difficult or impossible tasks. However, such bar configurations (fast rotation and long major semi-axis) are quite rare for Milky Way Analogue galaxies~\citep{Garma}.

\subsubsection{Resonance with Motions Perpendicular to the Disk Direction }

\begin{figure}[H]
\includegraphics[width=.39\textwidth]{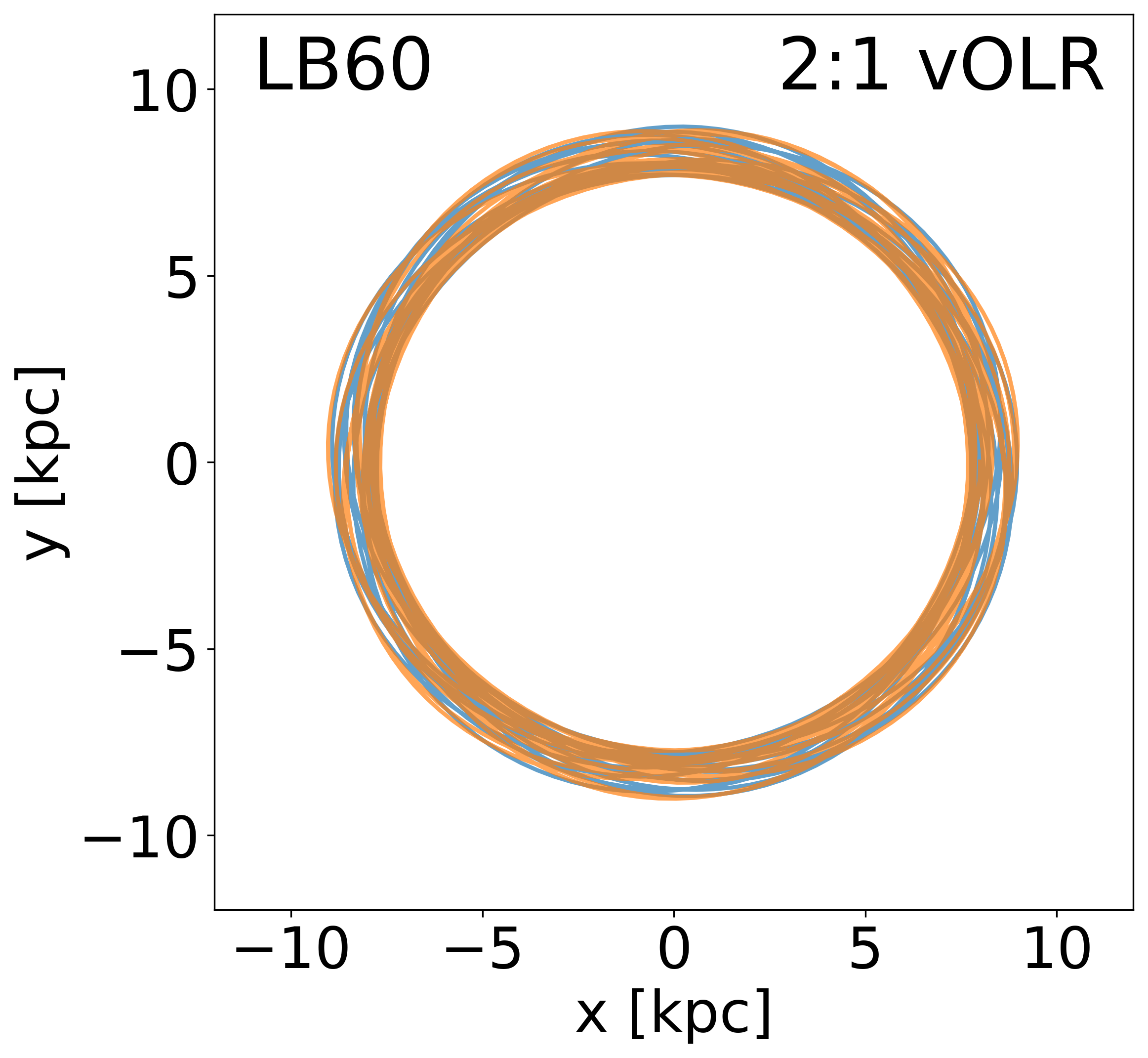}
\includegraphics[width=.42\textwidth]{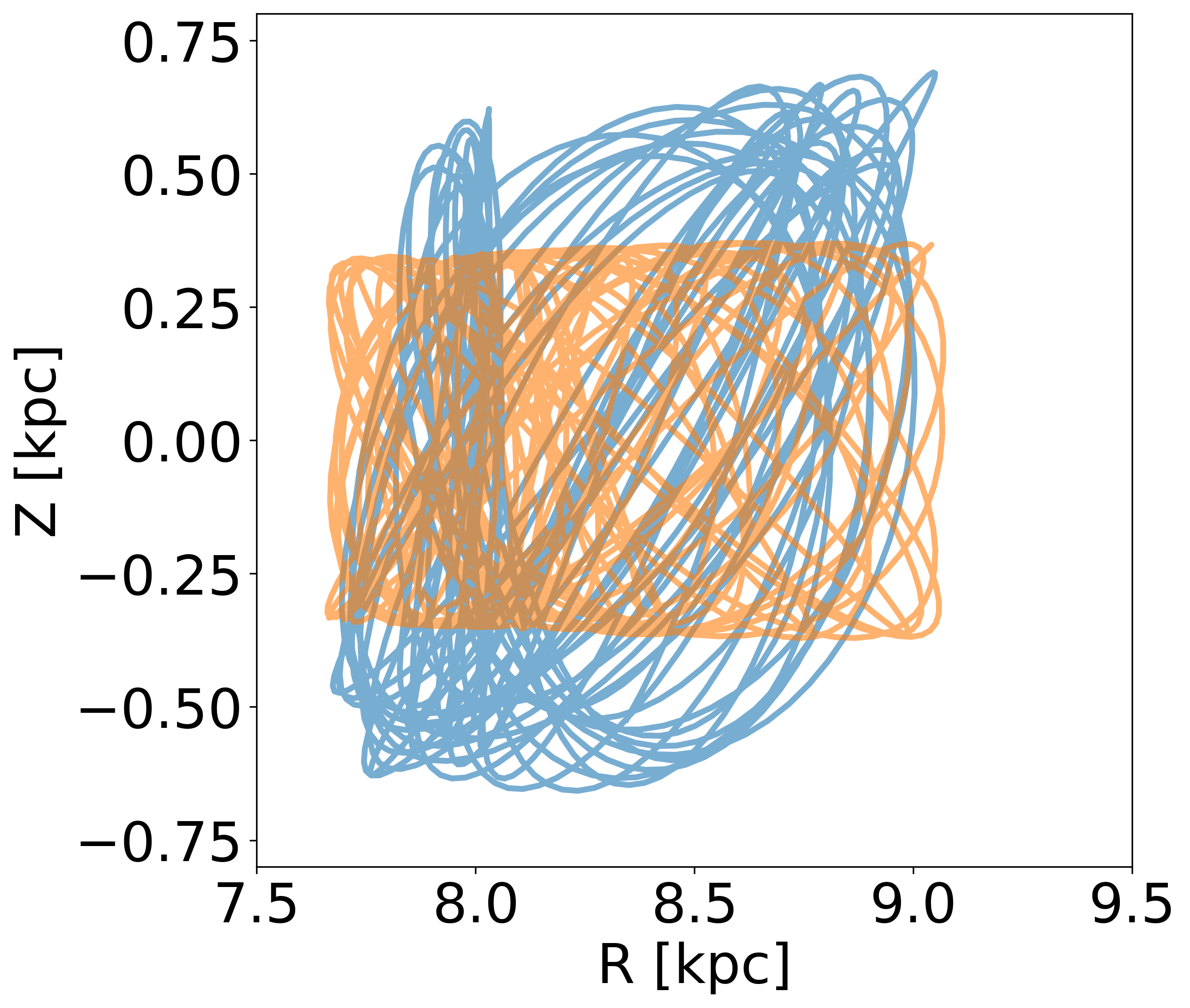}\\
\includegraphics[width=.42\textwidth]{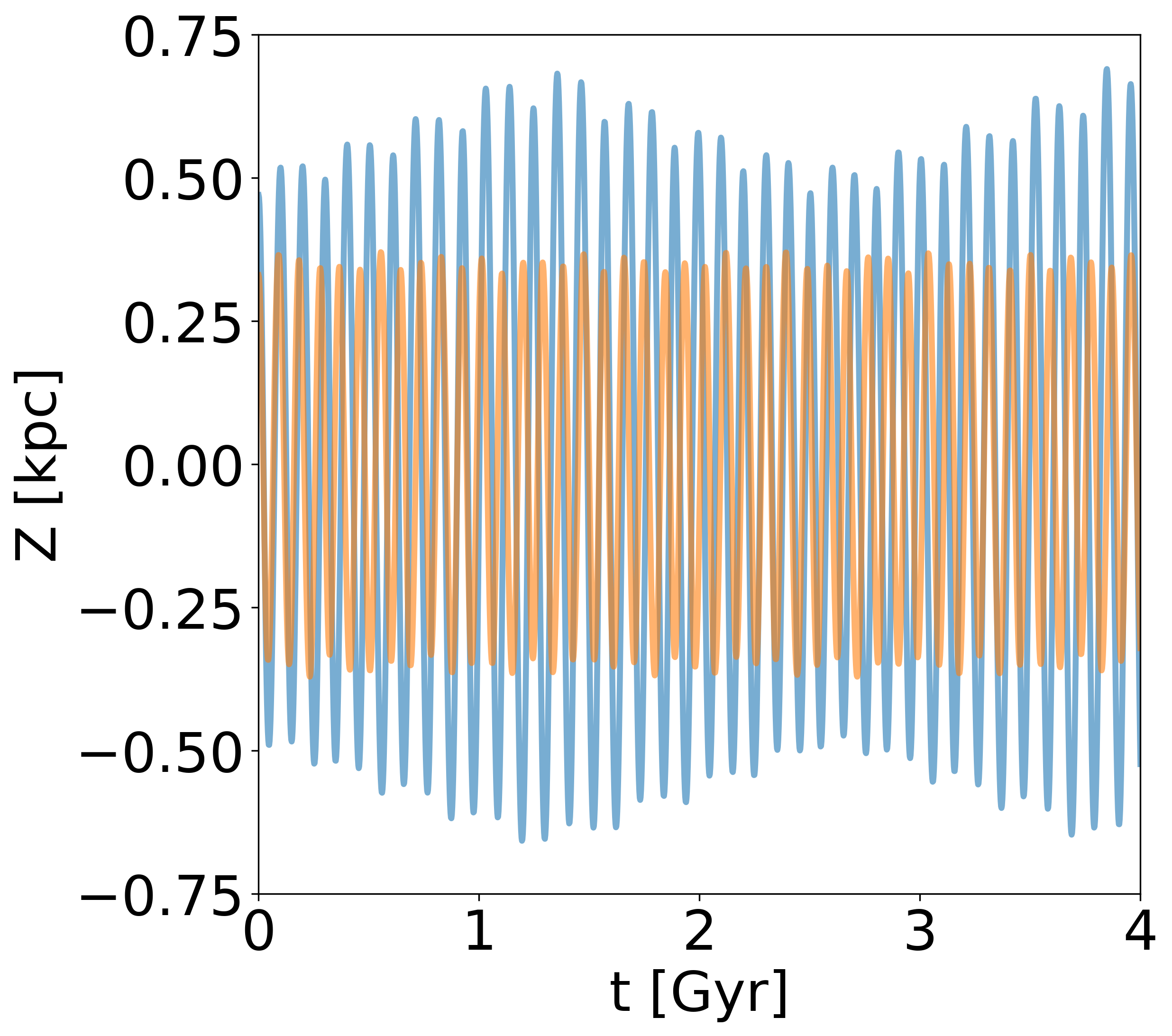}
\includegraphics[width=.40\textwidth]{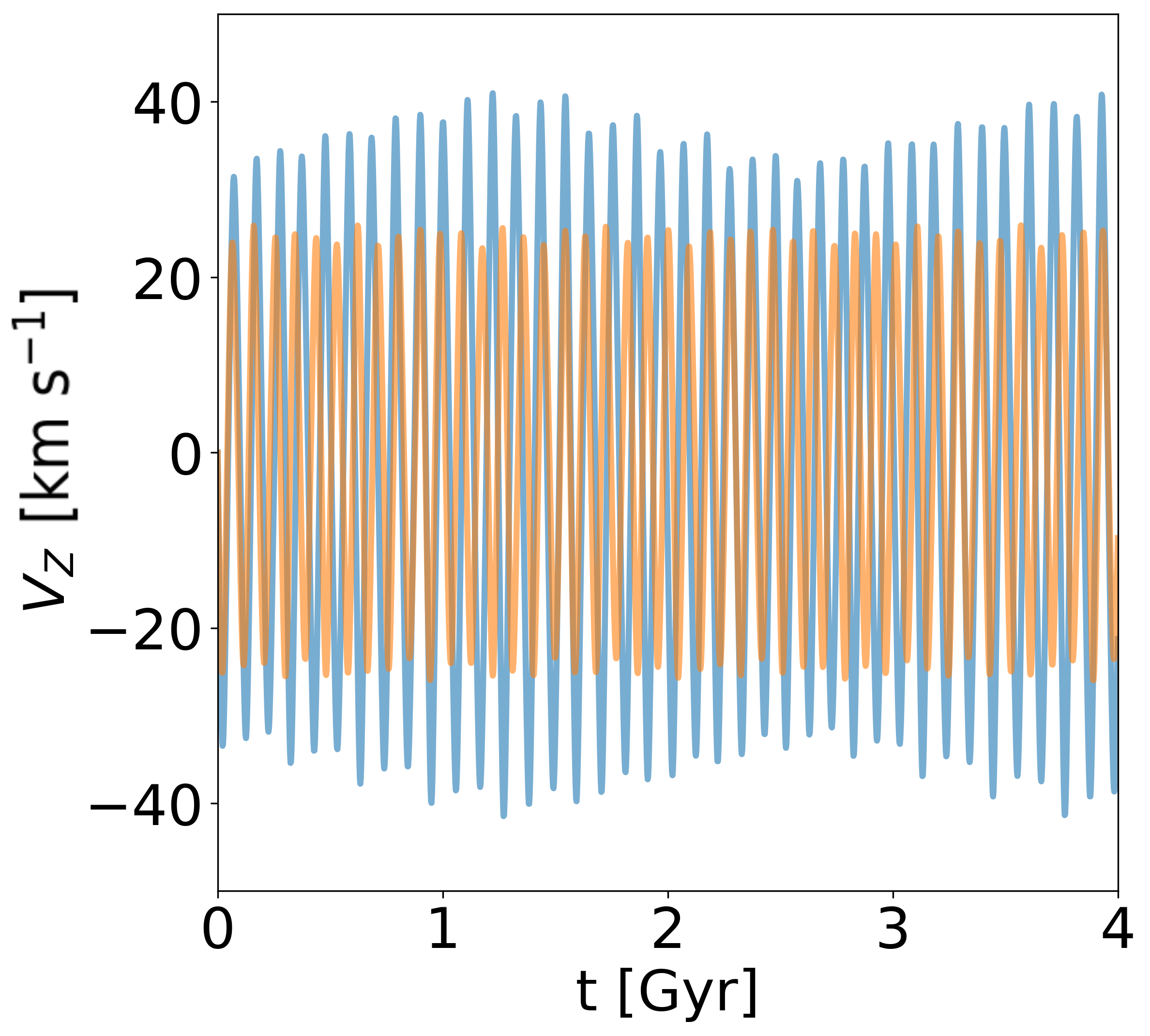}
\caption{Dynamics of resonant (blue) and non-resonant (orange) particles shown in the model LB60 on $(x,y)$, $(R,Z)$, $(t,Z)$ and $(t,V_Z)$ planes at 2:1 vOLR.   \label{volrs}} 
\end{figure}

Let us discuss the influence of the 2:1 vertical outer Lindblad resonance occurring
in the solar neighborhood on the disk's vertical structure if the angular velocity of the bar is equal to 60 $~\mbox{km}~\mbox{s}^{-1}~\mbox{kpc}$. 
For resonant buildup of oscillations, it is necessary that the azimuthal variation of the vertical force from the perturbing potential of a bar be considerable. We find that the model LB60, with its fast rotating and elongated bar, satisfies this criterion. To demonstrate the influence of 2:1 vOLR, we follow the dynamics of two test particles in the resonance region,  and integrate their orbits in the model LB60  with the fast rotating elongated bar.
These particles have initially circular orbits with initial $R=8~\mbox{kpc}$, but differ by frequencies of oscillations in the direction
perpendicular to the plane of the disk. The first particle has a frequency of oscillations equal to $\nu_{Z}\approx9.7~\mbox{Gyr}^{-1}$,  corresponding to the resonant frequency of oscillations at eight kpc (see Figure~\ref{figres}b). Another particle has a non-resonant frequency  perpendicular to the disk direction equal to $\nu_{Z}\approx12~\mbox{Gyr}^{-1}$. Figure~\ref{volrs} demonstrates the dynamics of the resonant and the non-resonant particles. As can be seen from the figure, in the plane of the disk, both particles have close to circular orbits, and oscillate along the radius with approximately the same amplitude.
However, the particles demonstrate different behavior when perpendicular to the disk direction.
Non-resonant particles (orange line) oscillate with constant amplitudes in both the coordinate  $Z$ and velocity $V_Z$. The resonant particles (blue line) oscillate perpendicular to the disk with periodically increasing and decreasing amplitudes of oscillations of $Z$ and $V_Z$ values. 

It should be noticed that the resonance with the particle motions perpendicular to the plane of the disk occurs at eight kpc 
if the velocities of the particles in that direction are about $V_Z\approx33~\mbox{km}~\mbox{s}^{-1}$ at Z = 0.  
Taking into account that the velocity dispersion  of the thin disk in the solar neighborhood is equal to $\sigma_Z=11~\mbox{km}~\mbox{s}^{-1}$, 
there is a small number of such stars. Moreover, we find that 2:1 vOLR resonance occurs if stars are in nearly circular orbits. 
Also, the build up of resonance oscillations depends on the phase of a star's motion perpendicular to the disk relative 
to the phase of a rotating bar.
Taking these factors  into account, we conclude that it is unfeasible to observationally confirm  a
manifestation of the 2:1 vertical resonance in the solar neighborhood.


\section{Summary}\label{sec:conclusion}
The results of our study can be summarized as follows.

In agreement with previous studies, we confirm that 2:1 OLR resonance of particles orbiting in the plane of the disk with the 
rotating bar leads to their spatial redistribution. The angular momenta of the resonant stars change under the influence of the bar. 
The maximum of such momentum exchange occurs at 2:1 outer Lindblad resonance. 
The momentum exchange has oscillational behavior when atop of the smaller 
time-scale oscillations corresponding to the periodic disturbances from the rotating bar, i.e., there are oscillations with a larger period. 

Spatial redistribution of the particles caused by their interaction with the potential of a bar is more prominent 
for a larger bar's semi-axis and its angular velocity is larger due to the fact that resonances in this case are closer to the perturbing potential. 
In addition to the redistribution of particles at OLR we find---in agreement with previous studies \linebreak  (\citet{Antoja2018})---that 
resonance leads to the formation of the ridges seen in $(R,V_{\phi})$ particles distribution.

The 2:1 and 1:1 OLR do not cause any noticeable influence on the distribution of particles and their velocities when perpendicular to the plane of the disk direction as a function of the radius. 
Such influence is seen, however, in the distributions of the average coordinate of the particles,$<|Z|>$ and their average velocity components,$<|V_Z|>$, as a function of the angular momentum $L_Z$. The effect manifests itself as bumps on $L_Z$---$<|Z|>$ and $L_Z$--$<|V_Z|>$- distributions developing outside the resonance region.

Simulations show an appearance of prominent bump of the average coordinate of the particles,$<|Z|>$ as a function of the momentum $L_Z$ at $\approx$  2200--2800 $~\mbox{km}~\mbox{s}^{-1}~\mbox{kpc}$ in the model with a long, fast bar. Appearance of the bump on $(-L_Z,<|Z|>)$---diagram is related to the 'escapers'---the particles at  2:1 OLR region that have the amplitudes of oscillations grown large enough so they reach the bar region. Strong interaction with the bar potential of such resonant particles provides an additional ``kick'' from a bar. We find that a few percent of the particles in the vicinity of the 2:1 OLR can  closely approach the bar region and 
increase their angular momenta from $-L_Z\approx$ 1600--1800 to \linebreak  2200--3000 $~\mbox{km}~\mbox{s}^{-1}~\mbox{kpc}$, propagating, as a consequence,
at large distances both in radius and in the direction perpendicular to the disk.

The escaping stars in the model with a long fast bar generate and strengthen with time the high- and low-velocity ``tails'' seen on the $(R,V_{\phi})$---particle distribution.

Another manifestation of the 2:1 resonance in the LB60 model is the appearance of the chevron-like structures seen on the $R,V_R$ 
diagram. Such structures are observed in the Milky Way halo and assumed to be the result of an accretion of a massive satellite galaxy into the Milky Way. Our simulations demonstrate
that chevrons can also be formed by the particles escaping from the 2:1 outer Lindblad resonance due to a strong interaction with the fast long bar.

The resonance of the particle motions in the direction perpendicular to the plane of the disk with the rotating bar (vOLR--resonance) can also affect the disk's kinematical properties in the direction perpendicular to its plane. We find that the ``vertical resonance'' can amplify the amplitude of oscillations of resonance particles by about a hundred percent. The condition for the vertical resonance in the solar neighborhood is not, however,  satisfied for all particles. Only particles, that have at the solar neighborhood the velocity of  $V_Z\approx33~\mbox{km s}^{-1}$ at Z = 0 satisfy the resonance condition. Moreover, we find that 2:1 vOLR resonance occurs if stars have nearly circular orbits. In addition, the build-up of resonance oscillations depends on the phase of a star's motion perpendicular to the disk relative to the phase of a rotating bar. Taking these factors into account, we conclude that it is unfeasible to observationally confirm a manifestation of the 2:1 vertical resonance in the solar neighborhood. 

The resonances 2:1 OLR and vOLR do not have any noticeable effect on \linebreak  $V_Z$-distribution of stars in 
the solar neghborhood. We conclude therefore that  
peculiarities of $V_Z$-velocity distribution in the solar neighborhood reported by \citet{Vieira} 
cannot be caused by the influence of resonances with a rotating bar.

\vspace{6pt} 



\authorcontributions{Conceptualization, V.K., G.C., K.V. and R.T.; methodology, V.K., G.C. and R.T.;
software, R.T. and A.L.; writing original draft preparation, V.K., R.T. and G.C.; writing review and editing, V.K., R.T., A.L., G.C. and K.V. All authors have read and agreed to the published version of the manuscript.}

\funding{The research by V.K. and R.T. was carried out at Southern Federal University with the financial support of the Ministry of Science and Higher Education of the Russian Federation (State contract GZ0110/23-10-IF).}



\dataavailability{All data used in this paper are taken from the open sources and the
references are given.} 

\acknowledgments {We thank the anonymous reviewers for the careful reading of the manuscript and valuable comments.} 

\conflictsofinterest{The authors declare no conflict of interest.} 


\abbreviations{Abbreviations}{
The following abbreviations are used in this manuscript:\\

\noindent 
\begin{tabular}{@{}ll}
Corotation Resonance & CR \\
inner Lindblad resonance & ILR \\
outer Lindblad resonance & OLR \\
vertical outer Lindblad resonance & vOLR \\
Long bar & LB \\
Short bar & SB \\
MC17 & McMillan17 \\
\end{tabular}
}


\begin{adjustwidth}{-\extralength}{0cm}

\reftitle{References}

\PublishersNote{}
\end{adjustwidth}
\end{document}